\documentclass{JHEP3} % 10pt is ignored!

\usepackage{epsfig}
\usepackage{graphicx}
\usepackage{wrapfig}
\usepackage{amsmath}
\usepackage{yfonts}
\usepackage{subfigure}
\usepackage{bm}% bold math

\def\be{\begin{equation}}\def\ba{\begin{eqnarray}}
\def\ee{\end{equation}}\def\ea{\end{eqnarray}}
\def\ben{\begin{enumerate}}\def\bitem{\begin{itemize}}
\def\een{\end{enumerate}}\def\eitem{\end{itemize}}
\def\no{\nonumber\\}

\def\calA{{\cal A}}\def\calB{{\cal B}}

\newcommand{\e}{{\mbox{e}}}

\def\roughly#1{\mathrel{\raise.3ex\hbox{$#1$\kern-.75em%
\lower1ex\hbox{$\sim$}}}}

\def\deriv{\partial}
\def\wn{\textswab{w}}
\def\qn{\textswab{q}}

\def\Im{{\textrm{Im}}} 
\newcommand{\dd}{{\rm d}}

\title{Photo-emission rate of sQGP at finite density}

\author{Kwanghyun Jo \\
Department of Physics Hanyang University, Seoul 133-791, Korea
\\ E-mail: \email{jokh38@gmail.com}}

\author{Sang-Jin Sin \\
Department of Physics Hanyang University, Seoul 133-791, Korea
\\ E-mail: \email{sjsin@hanyang.ac.kr}}

\received{\today}       %%
\abstract{We calculate the thermal spectral function of SYM plasma with finite density using holographic technique. The gravity dual of the finite temperature and density is taken as the RN-AdS black hole. In the presence of charge, linearized vector modes of gravitational and electromagnetic perturbation are coupled with each other. By introducing master variables for these modes, we solve the coupled system and calculate spectral function. We also calculate photo-emission rate of our gauge theory plasma from spectral function for light like momentum. AC, dc conductivity and their density dependence is also computed. }

\begin{document}

\section{Introduction}
The gauge/gravity duality \cite{Maldacena:1997re,Gubser:1998bc,Witten:1998qj} opened a new possibility to quantitative study for strongly interacting system. Although it is not  developed enough to describe realistic QCD, we expect to learn some features of QCD from it based on the universality
of the hydrodynamics: in longwavelenth limit, the details of the theory does not matter. For example,  the shear viscosity/entropy ratio \cite{Policastro:2001yc,Kovtun:2004de}  $\eta/s$ is universal if we neglect the higher derivative terms. We also expect, due to analytic structure of the theories, that there are  similarities of supersymmetric and non-supersymmetric theories can continue to the finite wavelenth/frequency regime.

 %In this sense, by using gauge/gravity correspondence, we compute some hydrodynamic quantities of boundary gauge theory from the dual gravity theory. Even though it is not the exact dual of real QCD, we believe that the result have some at least qualitative agreements with QCD.

The quarks and gluons are liberated  at high enough  temperature.
However, over $T_c < T <2\sim 3 T_c$, the experimental data shows that quarks and gluons are not  free but are strongly interacting: the small viscosity and the presence of the coherent flow show that the interactions should be very strong. Such strongness of the interaction is  the motivation why one has  to abandon perturbative QCD in such energy/temperature regime. One way to avoid that difficulty is to rely on hQCD for the quark gluon plasma in RHIC(Relativistic Heavy Ion Collider). The hydrodynamic calculations of hQCD were shown to be useful to discuss the transport phenomena \cite{Son:2002sd,Policastro:2002se}. It is  interesting to see what happens in LHC(Large Hadron Collider) where  the energy scale  is  much higher \cite{Abreu:2007kv}.

The finite baryon density effect is very essential ingredient to understand how the core of neutron star and the early universe behave. It maybe uncover some significant features of the evolution of our universe, galaxies and stars. At RHIC experiment, the temperature  reached is above $T_c$ but the density is almost zero. In near future J-PARC(Japan Proton Accelerator Research Complex), LHC and especially FAIR (Facility For Antiproton and Ion Research) which probes the regime of a few times of normal nuclear density \cite{Staszel:2010zz} will tell us  much about the density effect of quarks and gluons. The holographic study for the system with finite density in hydrodynamic regime  was made  in \cite{Son:2006em,Mas:2006dy,Maeda:2006by,RNtvmode,RNsound}.

The dual gravity background for the finite density and temperature is taken to be Reissner-Nordstr\"{o}m AdS blackhole. In the phase diagram, we know much about high temperature, low density regime but not low temperature high density regime. Previously we studied some issues ie. meson mass shifted by density effect for zero temperature, finite density sector \cite{Jo:2009xr}. Now we will study the effect of both finite temperature and density. The bulk U(1) charge is identified with the particle number density of the boundary field theory. To see the finite frequency/momentum  dependence of the response of the system, the spectral function is a good tool. It gives us ac conductivity and its trace is related to the photo-emission and di-lepton production rate \cite{CaronHuot:2006te,Parnachev:2006ev,Atmaja:2008mt}.

The spectral function with and without  baryon density in  the probe brane approach was already calculated  \cite{Kovtun:2005ev,Kovst,Teaney1,Myers:2007we,Erdm1,Mas:2008jz,Kaminski:2009dh}.  However in that approach the
gravity back reaction to the presence of the charge is neglected.
In this paper, we take bottom up approach where the back reaction is taked into account. We will compute  the spectral function of tensor and vector modes which describe the fluctuation of energy momentum tensor and currents of hot plasma. After that the   finite temperature and density effects of photo-emission rate are calculated and discussed.

\section{A recipe  for Greens function} \label{recipeforgreenf}
In this section, we will briefly review how to calculate thermal spectral function. To describe thermalized plasma holographically we need the black hole background. The general equations of motions for the linearized fluctuations in this background are
\be \label{eom1}
E_\alpha ''(u) + P(\wn,\qn,u) E_\alpha '(u) + Q(\wn,\qn,u) E_\alpha(u) =0
\ee
where $E_\alpha$ denote the fluctuating fields in given background and $\alpha$ runs 1 to n, the number of independent fields and $\wn, \qn$ are dimensionless frequency and momentum. Near the boundary,  ($u\sim 0)$, there are two local Frobenius solutions $\Phi_1, \Phi_2$
\ba \label{Frobeniussolnearbdry}
\Phi_1 &=& u^{\Delta_-}(1 + \cdots) \no
\Phi_2 &=& u^{\Delta_+} (1+ \cdots)
\ea
$\Delta_\pm$ is the solution of indicial equation near the boundary, $\Delta_+ > \Delta_-$ where $\Delta_+$ is the conformal dimension of an operator and $\Delta_-$ is the dimension of the dual source field. Near  the horizon,  u=1,  there are also two local solutions
\ba
\phi_1 & = & (1-u)^{-i \textswab{w}/2} (1+ \cdots) \no
\phi_2 & = & (1-u)^{i \textswab{w}/2} (1+ \cdots).
\ea
The two different local solutions of eq.(\ref{eom1}) should be matched:
\be
E_\alpha = \mathcal{A}(\wn,\qn)\Phi_1 +\mathcal{B}(\wn,\qn)\Phi_2 = \mathcal{C}(\wn,\qn)\phi_1 +\mathcal{D}(\wn,\qn)\phi_2.
\ee
However not all solutions are allowed physically because this system contains the blackhole: no outgoing-wave can propagate from the horizon, therefore we should impose , $\mathcal{D}=0$,
which is so called the infalling boundary condition. Taking the  normalization $\mathcal{C}=1$  using the linearity of differential equation, we have
\be \label{sol11}
\phi_1(u) = \mathcal{A}(\wn,\qn) \Phi_1(u) +\mathcal{B}(\wn,\qn) \Phi_2(u).
\ee
Note that the coefficient $\calA$ is the source of the boundary theory operator $\calA(\wn,\qn) = J_\alpha(\wn,\qn)$, so by differentiating the generating functional twice with respect to $J_\alpha$ we get the retarded Green function. And another coefficient $\calB$ corresponds to the condensate or vacuum expectation value of the operator ${\mathcal{O}}_\alpha$ which couples to the source $J_\alpha$. The retarded Green function is given by the ratio between $\mathcal{A, B}$. We will give a sketch of the on-shell quadratic action, with eq. (\ref{Frobeniussolnearbdry}), (\ref{sol11})
\ba
S_{\mbox{on-shell}} & \sim & \int d^{d+1}x \sqrt{g} {E^\alpha}' {E_\alpha}' \no
& \sim & \int d^{d} x \left[u^{-(\Delta_+ + \Delta_- -1)} \calA^2 \left(u^{\Delta_-}+\frac{\calB}{\calA}u^{\Delta_+}\right)\left(\Delta_- u^{\Delta_- -1}+\Delta_+\frac{\calB}{\calA}u^{\Delta_+ -1}\right)\right]_{u \rightarrow 0}\no
& \sim & \int d^{d} x \calA^2 \left[u^{-(\Delta_+ + \Delta_- -1)} \left(\Delta_- u^{2\Delta_- -1} + (\Delta_+ + \Delta_-)\frac{\calB}{\calA}u^{\Delta_+ + \Delta_- -1}\right)\right]_{u \rightarrow 0}\no
& \sim & \int d^{d} x \calA^2 \left[\Delta_- u^{\Delta_- -\Delta_+} + (\Delta_+ + \Delta_-)\frac{\calB}{\calA}\right]_{u \rightarrow 0}
\ea
obviously the first term in last line is divergent ($\Delta_- < \Delta_+$) so it should be renormalized holographically \cite{Skenderis:2002wp} or we can ignore it because the imaginary part of the Greens function do not care about the real number which comes the first term. For the issues how to regulate the on-shell action, see the appendix \ref{bdryaction}. The spectral function is its imaginary part \cite{Kovst}. Notice that
\ba \label{greensfunc}
\frac{\mathcal{B}}{\mathcal{A}} & = & \frac{1}{\Phi_2(u)} \left(\frac{\phi_1(u)}{\phi_1(0)}-\Phi_1(u) \right) \no
\chi &=& \Im G^{\mbox{ret}} = \Im \frac{\mathcal{B}}{\mathcal{A}} =  \frac{1}{\Phi_2}\Im \left(\frac{\phi_1(u)}{\phi_1(0)}\right), \quad \mbox{where}~ \mathcal{A} = \lim_{u\rightarrow 0}\frac{\phi_1 (u)}{u^{\Delta_-}}
\ea
Here $\Phi_i$ is real because the equation of motion and initial conditions are real for $\Phi_i$. Above expression is independent of $u$ since it is a kind of the conserved flux \cite{Son:2002sd}
\be
S_{bdry}[\phi_0]= \int \frac{d^4k}{(2\pi)^4} \phi_0(-k) \mathcal{G} (k,u) \phi_0(k) \bigg|^{u=1}_{u=0}
\ee
where $\mathcal{G} =\mathcal{N} \frac{\calB}{\calA}$, with $\mathcal{N}$ the normalization constant. The retarded Green function is defined by the recipe \cite{Son:2002sd}
\ba
G^R_{ij}(K) &=& -2 \mathcal{G}_{ij} (K,u=0) \quad i=j \no
&=& -\mathcal{G}_{ij} (K,u=0) \quad i \neq j
\ea
From the eq. (\ref{greensfunc}), the only thing we should know is the value of $\phi_1(u), \Phi_2(u)$. The spectral function, for example, is given when $\Delta_-$=0
\be
\chi_{ii} = -2 \mathcal{N} \lim_{u \rightarrow .5}\left[\frac{1}{\Phi_2(u)} \Im\left(\frac{\phi_1(u)}{\phi_1(u=0)}\right)\right].
\ee
Given  the normalization constant $\mathcal{N}$ we can calculate the spectral function numerically.

\section{RN AdS}
The dual geometry for the finite temperature and density is chosen as charged AdS black hole \cite{RNsound}. The  action is
\be \label{RNaction1}
S = \frac{1}{2 G_5^2} \int d^5 x \sqrt{-g} ({\mathcal{R}}-2\Lambda)+\frac{1}{4g_5^2} \int d^5 x \sqrt{-g}~F^2 +\frac{1}{G_5^2}\int d^4 x \sqrt{-g^{(4)}}K,
\ee
where the cosmological constant is $\Lambda = -\frac{(d-1)(d-2)}{2 l^2}$, the last term is the Gibbons-Hawking term. And K is the extrinsic curvature on the boundary, l is the AdS radius. The metric of RN AdS is
\ba
ds^2 &=& \frac{r^2}{l^2} \left(-f(r) dt^2+d\vec{x}^2\right)+\frac{l^2}{r^2 f(r)}dr^2 \no
f(r) &=& 1-\frac{m l^2}{r^4}+\frac{q^2 l^2}{r^6}, \quad A_t=-\frac{Q}{r^2} + \mu
\ea
where the gauge charge Q is related to the black hole charge q
\be
g_5^2 = \frac{2 Q^2}{3 q^2} G_5^2, \quad Q^2 = \frac{3 g_5^2}{2 G_5^2} q^2,
\ee
the five dimensional gauge theory coupling constant $g_5$ and the gravitational constant $G_5$ \footnote{Usually five dimensional gravitational constant $G_5$ is used as $\kappa^2 = 8\pi G_5$ but here we will use $G_5$ as $\kappa$ itself.} can be chosen as \cite{SSJBFB}
\be
\frac{l}{g_5^2} = \frac{N_c N_f}{4 \pi^2}, \quad \frac{l^3}{G_5^2} = \frac{N_c^2}{4\pi^2}
\ee
but we will not use these parameters explicitly. The metric function $f(r)$ is rewritten as
\be
f(r) = \frac{1}{r^6} (r^2 - r_+^2)(r^2 - r_-^2)(r^2 - r_0^2), \quad
r_\alpha^{-2} = \frac{m}{3q^2} \left(1+ 2 ~ \mbox{cos}\left(\frac{\theta}{3}+ \varphi_\alpha \right) \right)
\ee
where $\alpha = +, -, 0$ and $\varphi_+ = 4 \pi/3, ~ \varphi_- = 0, ~ \varphi_0 = 2 \pi/3$.
The charge is expressed by $\theta$ and m,
\be \label{bhpararelation}
q^4 = \frac{4 m^3 l^2}{27} \mbox{sin}^2 \left(\frac{\theta}{2}\right), \quad
\theta = \mbox{arctan} \left( \frac{3 \sqrt{3} q^2 \sqrt{4m^3l^2-27q^4}}{2m^3l^2 - 27q^4} \right)
\ee
the range of outer horizon is $\sqrt{\frac{m}{3}} l \leq r_+^2 \leq \sqrt{m} l$.
Finally, black hole temperature is given as
\be
T = \frac{r_+^2 f'(r_+)}{4 \pi l^2}=\frac{r_+}{\pi l^2}\left(1-\frac{q^2l^2}{2 r_+^6}\right) \equiv
\frac{1}{2\pi b} \left(1-\frac{a}{2} \right)
\ee
where $a, b$ are
\be
a \equiv \frac{q^2 l^2}{r_+^6}, \quad b \equiv \frac{l^2}{2 r_+} .
\ee
From the horizon regularity, the black hole charge q is related with the chemical potential $\mu$
\ba
\mu &=& \frac{4 Q b^2}{l^4} = \frac{1}{2b}\frac{g_5 l}{G_5}\sqrt{\frac{3a}{2}}, \no
a&=& \frac{3+2\bar{\mu}^2 - \sqrt{9+12\bar{\mu}^2}}{\bar{\mu}^2}, \quad \mbox{where}~ \bar{\mu}=\frac{\mu}{T}\frac{G_5}{\pi g_5 l}.
\ea
Notice that there is maximum value of q, $q^4 \leq \frac{4}{27}m^3 l^2$ which corresponds to $a=2$. Horizon radius $r_+$ should be real, so 1+2 cos($\theta/3+4\pi/3$) must be  positive.

It is very useful to express the frequency and momentum as a dimesionless quantities $\wn=w/(2\pi T), \qn=k/(2\pi T)$. This choice is good enough to see the finite temperature behavior of system but not good in zero temperature limit. Alternative way is to rescale w and k by the black hole radius $r_+$, that is   by $b$: let  $\tilde{\wn} = b w, \tilde{\qn}= b k$. At the extremal limit, by eq. (\ref{bhpararelation}), $4m^3l^2 = 27q^4$  and $\theta = \pi$
\be
r_+ = \left(\frac{l^2}{2}\right)^{1/6} q^{1/3}, \quad a= 2, \quad b = \left(\frac{l^2}{2}\right)^{5/6} q^{-1/3}
\ee
and the chemical potential is written
\be
\mu = \frac{\sqrt{3}}{2} \frac{e l}{G_5} \frac{1}{b}.
\ee
If we rescale w and k with b, in the extremal limit we rescale w and k with chemical potential, $\tilde{\wn} \sim w / \mu, \tilde{\qn} \sim k / \mu$.

The origin of charged black hole in string theory can be understood by STU model : the diagonally charged STU black hole is RN AdS.
% or Einstein DBI system in $\alpha '$ goes to zero limit
%$\alpha'$ expansion of DBI action is essentially same as $1/N_c$.   Fixing  the $N_f/N_c$ finite, Yang-Mills term is only survived and others are vanished. We believe that is related to the Veneziano limit of boundary field theory, so this geometry is dual to the beyond quenched approxiamation.
% %The physical meaning of  U(1) charge of the RN AdS blackhole.
The diagonal $U(1)$ is the subgroup of the SU(4) R-symmetry originally but here we assume that this $U(1)$ is a part of flavor U(1) group which is relevant if we assume that the bulk filling branes\cite{SSJBFB} are embedded in our AdS5 space time \footnote{Please note that this is no more than a conceptual introduction of bottom up approach in string theory.}. The merit of doing this is that we can have a back-reacted gravitational background which is a solution of glue-quark coupled system in terms of gauge theory it means our approach is beyond quenched approximation. 

%The U(1) subgroup of flavor symmetry is diagonal and it counts only numbers of quark, no matter what they are charged under flavor group. Therefore, even though $U(1)_R$ charge and $U(1)_B$ charge are different, they play same role in counting the number of fields. We will use this $U(1) $ charge as number density of fermions. And also we assume that we have fundamental Fermions which is introduced by the bulk filling brane. Since our original D3 geometry or AdS$_5$ is dual to super Yang-Mills which have adjoint Fermions not fundamental. The bulk filling brane, however, introduces the flavor degrees of freedoms which are fundamental Fermions so it gives a kind of analogy to the dense system. Or simply saying that our RN AdS$_5$ is chosen as a gravity dual of four dimensional gauge theory which has fundamental degrees of freedom with finite temperature and density.

\section{Tensor mode}
The gravitational and gauge field perturbation is classified by the boundary SO(2) rotational symmetry. This classification is summarized in the appendix  \ref{Appd1}. Tensor mode perturbation is easy to treat because it is completely decoupled from other fields. The equation of motion for $h_{xy}$ component is
\be \label{eomtensorwoscal}
{h^x_y}''+\frac{(r^5 f)'}{r^5 f} {h^x_y}' +\frac{l^4}{r^4 f^2}\bigg(w^2 - k^2 f \bigg) h^x_y = 0
\ee
where prime denotes derivative with respect to r and $h^x_y = g^{(0)xx} h_{xy}$. Introducing new coordinate
$u = r_+^2/r^2$,
\be
ds^2 = \frac{(\pi T l)^2}{u} \left(-f(u) dt^2+d\vec{x}^2\right)+\frac{l^2}{4 u^2 f(u)}du^2 ,\quad
f(u) = (1-u)(1+u-au^2).
\ee
The equation of motion is simplified
\be \label{eomtensorinu}
{h^x_y}''-\frac{f-u f'}{u f} {h^x_y}' +\frac{1}{u f^2} \bigg({\tilde\wn}^2 - {\tilde\qn}^2 f \bigg) h^x_y = 0.
\ee
 where % rescaled frequency and momentum are defined by $\wn = \frac{w}{2 \pi T}$, $\qn = \frac{k}{2 \pi T}$.
\be
\tilde{\qn}=b k = \frac{k}{2 \pi T}\left(1-\frac{a}{2}\right) = \qn \left(1-\frac{a}{2}\right), \quad
\tilde{\wn}= \frac{w}{2 \pi T}\left(1-\frac{a}{2}\right) = \wn \left(1-\frac{a}{2}\right).
\ee
This differential equation has two  independent solutions near boundary,
\be \label{soltens}
h^x_y = \calA \Phi_1(u) + \calB \Phi_2(u)
\ee
where
\be \label{localseries}
\Phi_1(u) = 1+ \cdots + h_l \mbox{log}(u) \Phi_2(u), \qquad \Phi_2(u) = u^2 (1+ \cdots).
\ee
In eq. (\ref{sol11}), we choose  $\calA = \phi_1(0)$. By dividing $\calA$ both side, we get the normalized solution for tensor mode
\be \
h^x_y(u) = \frac{\phi_1(u)}{\phi_1(0)} = \Phi_1(u) + \frac{\calB}{\calA} \Phi_2(u).
\ee
The on-shell action is given in eq. (5.7) of \cite{RNtvmode},
\be
S[h^x_y] = \frac{l^3}{32 G_5^2 b^4} \int \frac{d^4 k}{(2 \pi)^4} \left(\frac{f(u)}{u} h^x_y (-k, u){h^x_y}'(k,u)\right)\bigg|^{u=0}_{u=1}.
\ee
Our normalization is such that $h^x_y(u) \to 1$ at the boundary.
\footnote{More properly, we should express $h^x_y (u,k)= \hat{{h^x_y}}(k) h^x_y(u)$, where hatted variable is the value at the boundary or the external source of the boundary theory.}  By taking the imaginary part of the Greens function and renormalizing divergent terms, the thermal spectral function is
\be
\chi(\wn,\qn)_{xy,xy} = \frac{l^3}{16 G_5^2 b^4} \mbox{Im}\left(2\frac{\calB}{\calA}\right).
\ee
Here the ratio $\calB/\calA$ is
\be
\frac{\calB}{\calA} = \frac{1}{\Phi_2(u)}\left(\frac{\phi_1(u)}{\phi_1(0)} - \Phi_1(u)\right).
\ee
This ratio is independent of the evaluation point. As explained before, imposing the infalling condition at the   horizon and Dirichlete boundary condition at the UV boundary, we get the numerical solution for $\phi_1(u)$ and $\Phi_2(u)$,
\be
\chi(w,k)_{xy,xy} = \frac{l^3}{16 G_5^2 b^4} \mbox{Im}\left(
\frac{2}{\Phi_2(u)}\frac{\phi_1(u)}{\phi_1(0)}\right)
%= \frac{(2 \pi T)^4 l^3}{16 G_5^2 (1-a/2)^4} \mbox{Im}\left(\frac{2}{\Phi_2(u)}\frac{\phi_1(u)}{\phi_1(0)} \right).
\ee
Using $b=(1-a/2)/2\pi T$, one can show that
 the zero temperature spectral function is
\be
\chi(w,k)_{xy,xy}^{T\to 0}= \frac{(2 \pi T)^4 l^3}{16 G_5^2} \pi (\wn^2-\qn^2)^2 \theta(w^2-k^2).
\ee
See   appendix \ref{largewsp} for detail.
Figure \ref{tensorspectralqzero} shows the difference between normalized thermal spectral function at non-zero and zero temperatures for $\Delta\chi_{xy,xy}$. The thick line is zero chemical potential case $\bar{\mu}$=0 which is the  AdS Schwarzschild case of ref. \cite{Teaney1}. The dashed and solid line  correspond  to the $\bar{\mu}$ = 0.5, 1 case respectively. When  the chemical potential $ {\mu}$ increases, the spectral difference grows up and the position of the small peak is shifted to the larger $\wn$.
\begin{figure}
  \includegraphics[angle=0, width=0.45 \textwidth]{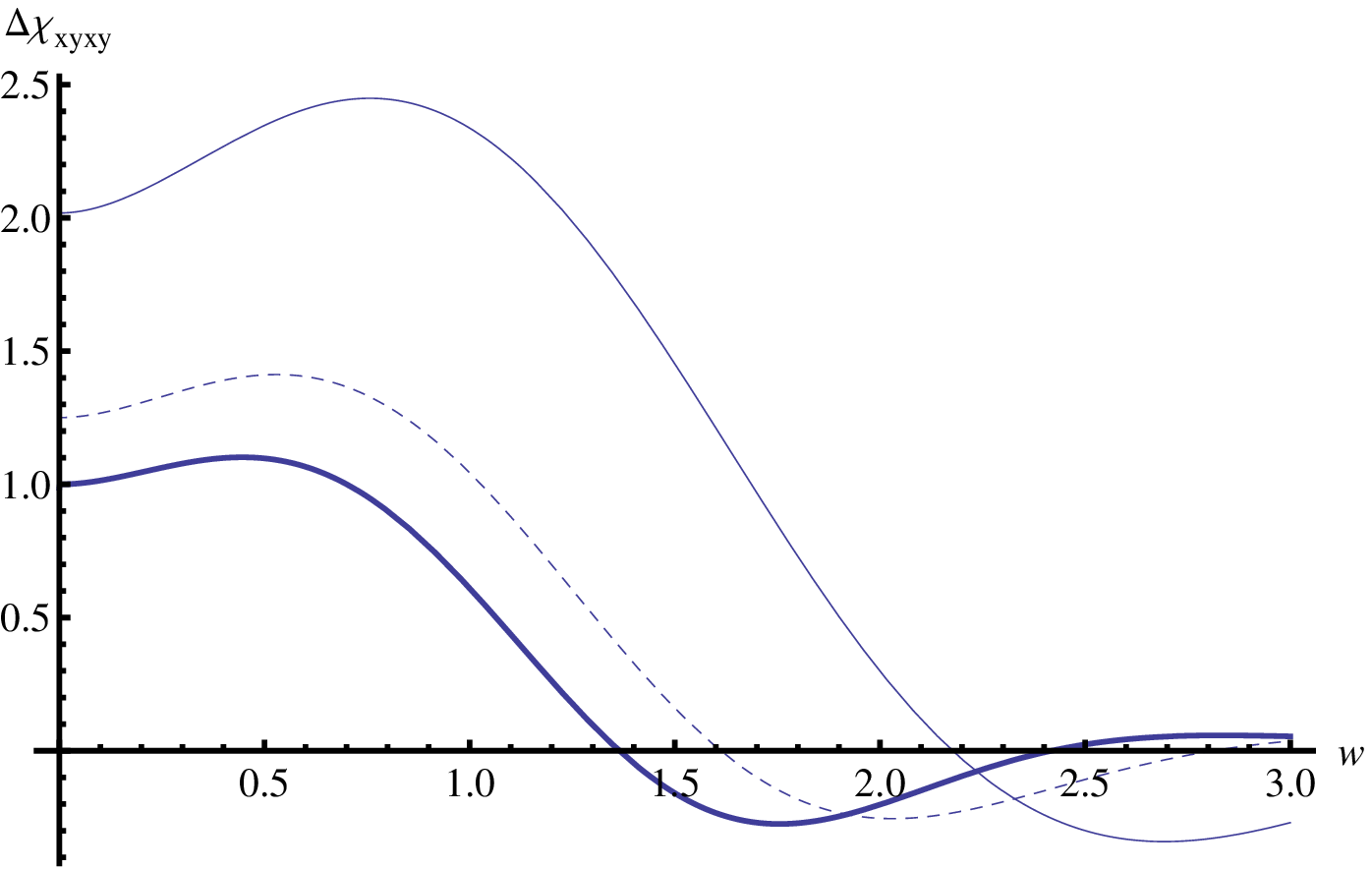}
  \includegraphics[angle=0, width=0.45 \textwidth]{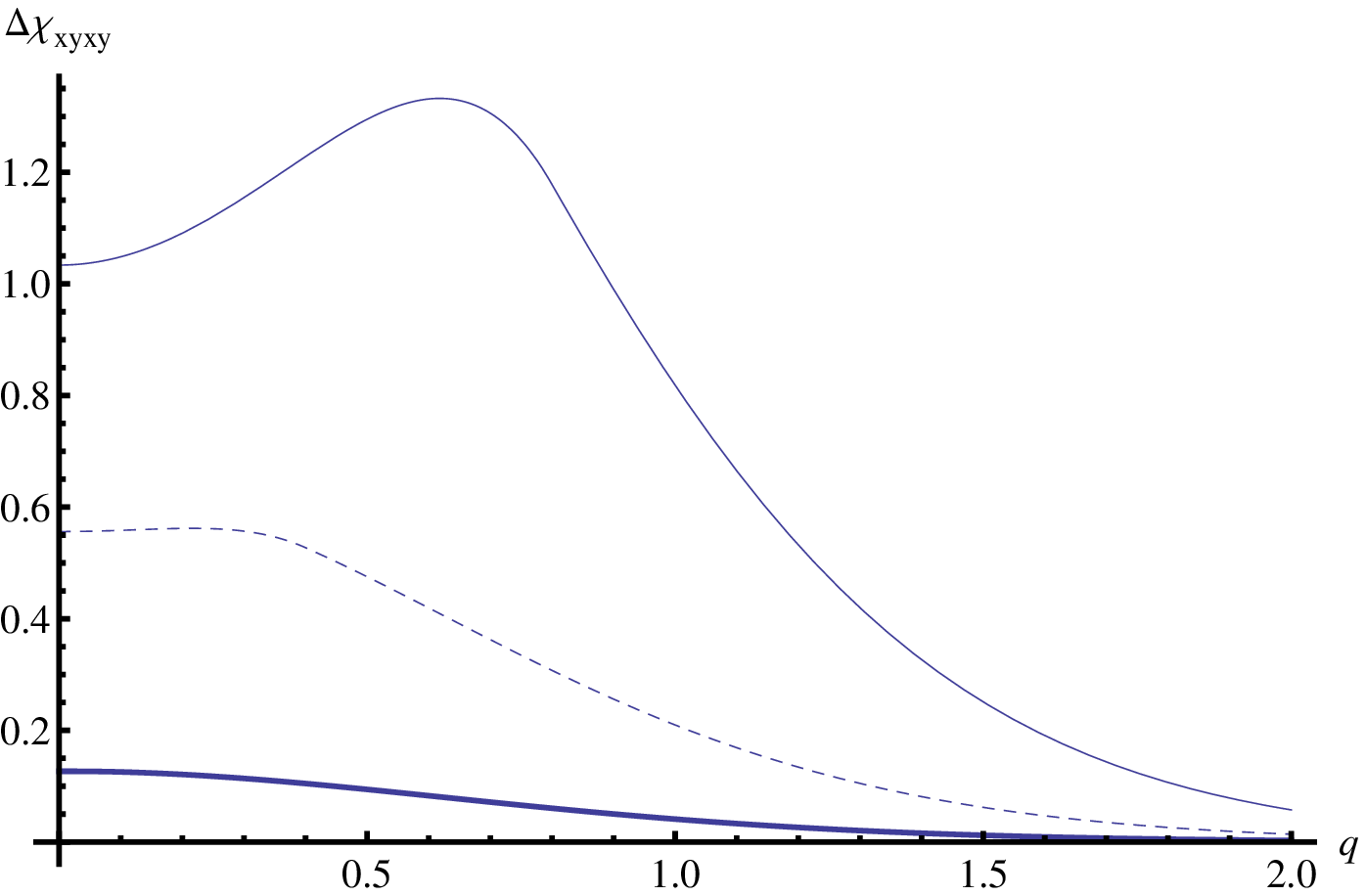}
  \caption{The difference of thermal spectral function and zero temperature one of $\Delta\chi_{xy,xy}$/$\wn$. Left: with fixed $\qn$ =0 when $\bar{\mu}$=0(thick), 0.5(dashed), 1(thin), right: $\Delta\chi_{xy,xy}$ vs $\qn$ plot and w=0.1(thick), 0.4(dashed), 0.8(thin) with fixed $\bar{\mu}$=0.5}   \label{tensorspectralqzero}
\end{figure}
The position of peak in the spectral difference is the pole position of retarded Greens function \cite{Kovtun:2005ev}. The shift of the peak is the shift of the quasi normal mode. When chemical potential grows in the unit of T, the pole position grows faster than T.

The right side of the fig. \ref{tensorspectralqzero} shows the spectral function as a function of spatial momentum $\qn$. When $\wn$=0, the peak position can be identified with the inverse screening length. However, for the tensor mode, there is no peak for $\wn$=0.
%It means for the static limit, super Yang-Mills plasma screening length goes to infinity so this plasma will be described as perfect liquid.
As we know SYM has conformal symmetry, hence it can not have any scale. For the finite $\mu$, there is a broad peak and the peak is more sharpened when chemical potential grows. This shows that in a dense system the thermal particle collides more often so that the particles propagates shorter distance. For the light like momentum, this screening is maximized.  For spacelike momentum, the thermal fluctuation of spectral function rapidly vanishes  leaving only zero temperature piece.
% spectral function remains.

\section{Vector mode}
\begin{figure}
\begin{center}
    \includegraphics[angle=0, width=0.48 \textwidth]{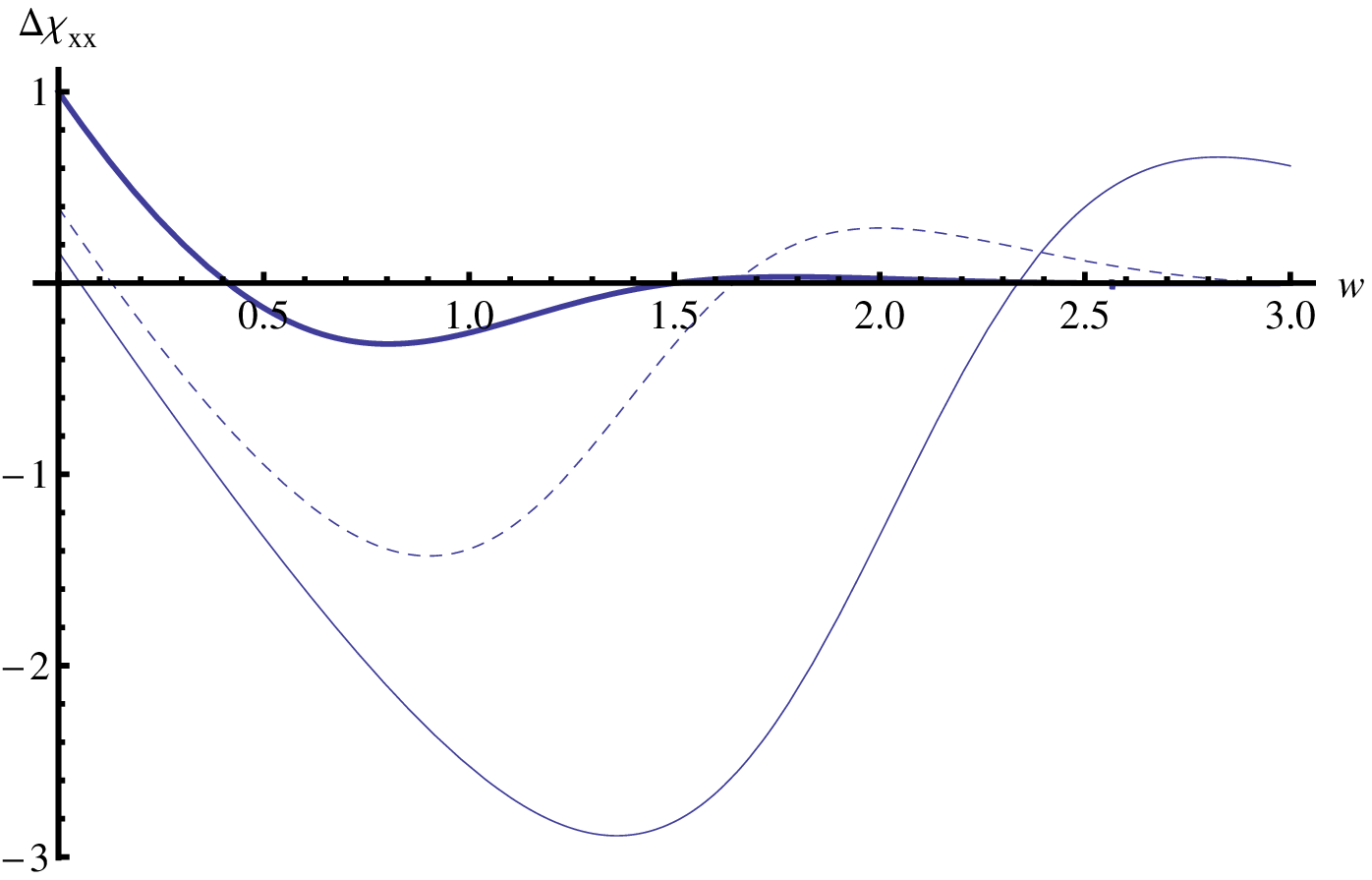}
    \includegraphics[angle=0, width=0.48 \textwidth]{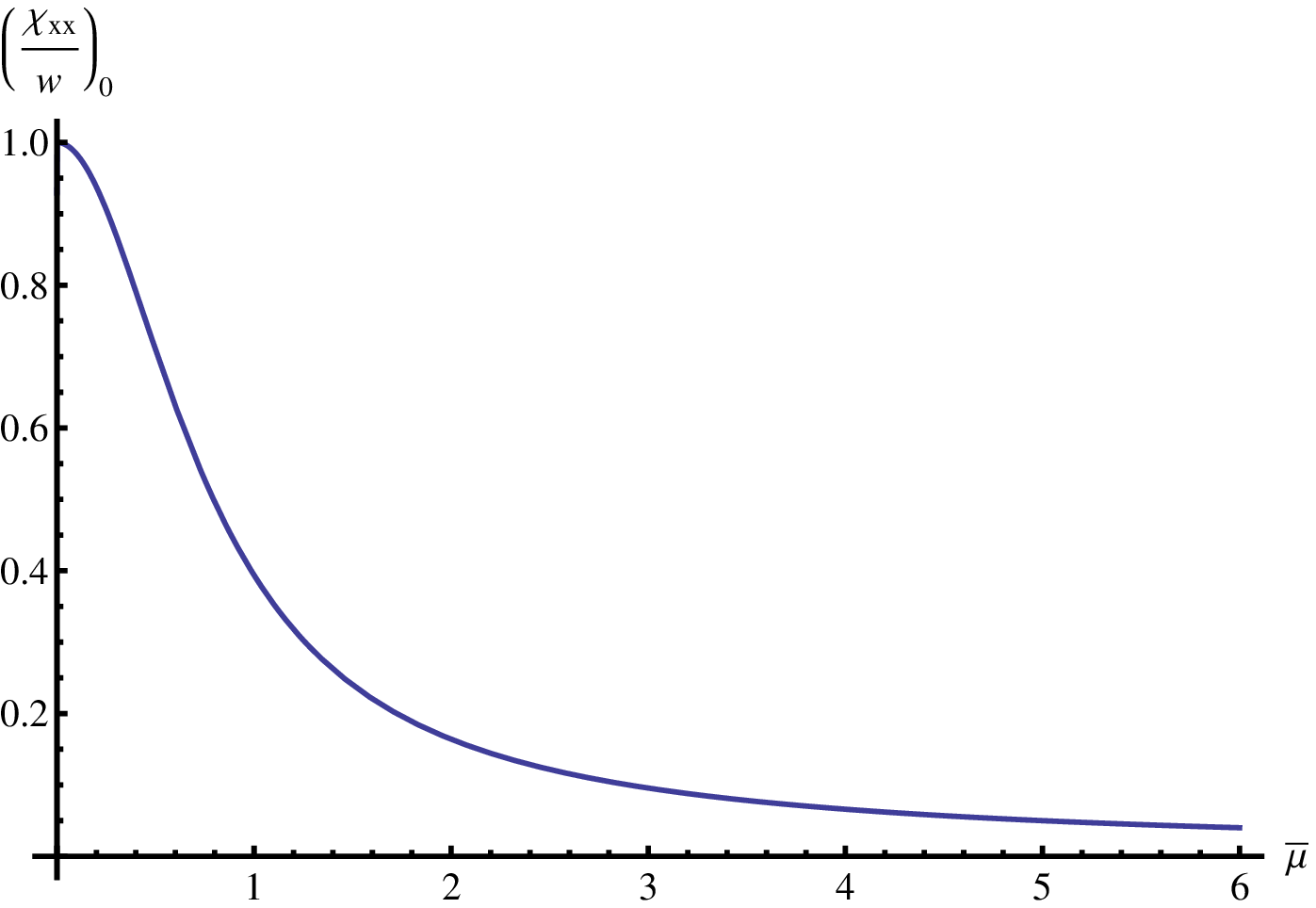}
  \caption{$\Delta\chi_{xx}$/$\wn$, Deviation of finite temperature  thermal spectral function from the zero temperature one,  with $\bar{\mu}$=0(thick), 1(dashed), 2. The normalization unit is $(2 \pi T)^2 \frac{l}{2e^2}$. Right : The density dependence of DC conductivity with normalization constant $\frac{1}{2}\frac{l}{e^2} 2\pi T$}\label{vsqnzero}
\end{center}
\end{figure}

Vector type perturbation consist of three independent fields, $h_{xt}, h_{xz}, A_x$ and equation of
motion for these modes are coupled with each other.
In hydrodynamic limit, this mode has a diffusion pole so that
it is also named as diffusive mode. Here we are interested in
general energy/momentum regime.
  The equations of motion for vector modes are
\ba \label{coupledeqs}
0 & = & {h^x_t}''-\frac{1}{u}{h^x_t}' -\frac{(1-\frac{a}{2})^2}{uf}(\wn\qn h^x_z + \qn^2 h^x_t) - 3a u B' \no
0 & = & \qn f {h^x_z}'+\wn{h^x_t}'- 3a \wn u B\no
0 & = & {h^x_z}'' + \frac{(f/u)'}{f/u}{h^x_z}' + \frac{(1-\frac{a}{2})^2}{u f^2}(\wn^2 {h^x_z}+\wn\qn h^x_t)\no
0 & = & B''+\frac{f'}{f}B' + \frac{(1-\frac{a}{2})^2}{uf^2}\left(\wn^2 -\qn^2 f\right)B -\frac{{h^x_t}'}{f} .
\ea
This equation is simplified by introducing gauge invariant combination $Z_1 = \wn h^x_z + \qn h^x_t$,
\ba
0 &=& Z_1''+\left(\frac{f' \tilde{\wn}^2}{f \left(\tilde{\wn}^2-\tilde{\qn}^2 f\right)}-\frac{1}{u}\right) Z_1' + \frac{\left(\tilde{\wn}^2-\tilde{\qn}^2 f\right)}{u f^2}Z_1 \no
&&-3 a u \tilde{\qn}\left(B'+\frac{\tilde{\wn}^2 f'}{f \left(\tilde{\wn}^2-\tilde{\qn}^2 f\right)} B \right)  \no
0 &=& B''+\frac{f'}{f}B' +\left[\frac{\tilde{\wn}^2-\tilde{\qn}^2 f}{u f^2}-\frac{3 a u}{f}-\frac{3 a u \tilde{\qn}^2}{\tilde{\wn}^2-\tilde{\qn}^2 f}\right]B+\frac{\tilde{\qn}}{(\tilde{\wn}^2-\tilde{\qn}^2 f)} Z_1'
\ea
where $\tilde{\wn}=\left(1-\frac{a}{2}\right)\wn$. It is not easy to solve these 2nd order coupled differential equations, but the authors of \cite{RNtvmode} have decoupled these equations by introducing master variables. Let us define first $\Psi_\pm$ as
\be
\Psi_\pm = -\frac{f\tilde{\qn}}{\tilde{\wn}^2-\tilde{\qn}^2 f} Z_1' +\left(-3au \frac{f\tilde{\qn}^2}{\tilde{\wn}^2-\tilde{\qn}^2 f}+C_\pm\right)B.
\ee
The equation of motion is rewritten as
\be
\Psi_\pm'' + \frac{f'}{f} \Psi_\pm' +\left(\frac{\tilde{\wn}^2-\tilde{\qn}^2 f}{uf^2}- \frac{f'}{uf}-\frac{C_\pm}{f} \right)\Psi_\pm=0,
\ee
where $C_\pm$ is
\ba
C_\pm &=& (1+a)\pm \sqrt{(1+a)^2+3ab^2 k^2} = (1+a)\pm \sqrt{(1+a)^2+3a \left(1-\frac{a}{2}\right)^2 \qn^2} \no
&=& (1+a)(1\pm\gamma), \quad \mbox{where} \quad  \gamma= \sqrt{1+\frac{3a}{4}\left(\frac{2-a}{1+a}\right)^2\qn^2}.
\ea
In order to use our recipe for spectral function,
we need   the on-shell action for vector modes:
\ba
S_{os} &=& \frac{l^3}{32 G_5^2 b^4} \int \frac{d^4 k}{(2 \pi)^4}
\bigg( \frac{1}{u} h^x_t (-k,u) {h^x_t}'(k,u) \no
&& - \frac{f(u)}{u}h^x_z(-k,u){h^x_z}'(k,u) -3a f(u) B(-k,u)B'(k,u)\bigg)\bigg|^{u=1}_{u=0}\no
&=& \frac{l^3}{32 G_5^2 b^4} \int \frac{d^4 k}{(2 \pi)^4}\bigg(\frac{1}{u} \frac{f}{\tilde{\wn}^2-\tilde{\qn}^2 f}Z_1 Z_1' -3afBB'+\frac{\tilde{\wn}^2}{\tilde{\wn}^2-\tilde{\qn}^2 f} 3au B h_t\bigg).
\ea
From the  equations for master field we can get the spectral function of master fields. We however need the spectral function of original variables not the master fields itself. In ref. \cite{Edalati:2010hk}, authors showed a systemetic way to compute the spectral function of original variables in terms of master variables. Let us first find the series solution of $Z_1, B$,
\ba
Z_1 &=& \hat{Z}_1(1+(\tilde{\wn}^2-\tilde{\qn}^2)u+\cdots) +\frac{\pi_Z}{2} u^2 + \cdots \no
B &=& \hat{B}(1+\cdots)+ \pi_B u  + \cdots \nonumber,
\ea
which defines the conjugate momentums $\pi_Z,\pi_B$.
The boundary action can be written in terms of  the boundary values of original variables and their conjugate momentums:
\be
S_{bd} = \frac{l^3}{32 G_5^2 b^4} \int \frac{d^4 k}{(2 \pi)^4}\bigg( \frac{\hat{Z}_1 \pi_Z }{\tilde{\wn}^2-\tilde{\qn}^2}-3a \hat{B} \pi_B + \cdots \bigg), \ee
where $\hat{Z}_1, \hat{B}$ are the boundary values of the fields and $\pi_Z ,\pi_B$ are their conjugate momentum which will be identified with one point function and dots denote contact terms which does not have any derivatives with respect to u. Note that these conjugate momentums depends on  the boundary source terms implictly. The master variables $\Psi_{\pm1}$ and $\Psi_{\pm 2}$ have series solutions near the boundary,
\be \label{masterseriessol}
\Psi_\pm = \hat{\Psi}_\pm(1+ \cdots + \hat{\Pi}_\pm u +\cdots) \no
\ee
Define the transformation matrix R,
\be
\mbox{R}= \left(
\begin{array}{cc}
-\frac{\tilde{\qn}}{\tilde{\wn}^2-\tilde{\qn}^2} & C_+ \\
-\frac{\tilde{\qn}}{\tilde{\wn}^2-\tilde{\qn}^2} & C_-
\end{array} \right),
\ee
the boundary value of the master fields is simply related to the boundary value of the original fields by R,
\ba
\left(
\begin{array}{c}
\Psi_+ \\ \Psi_-
\end{array} \right)&=&\mbox{R}
\left(
\begin{array}{c}
(Z_1)'(u=0)\\\hat{B}
\end{array} \right)
=\mbox{R} \left(
\begin{array}{c}
(\tilde{\wn}^2-\tilde{\qn}^2)\hat{Z}_1\\\hat{B}
\end{array} \right) \no
\left(
\begin{array}{c}
\Psi_+ \Pi_+ \\ \Psi_- \Pi_-
\end{array} \right)&=&\mbox{R}
\left(
\begin{array}{c}
\pi_Z \\ \pi_B
\end{array} \right)
\ea
Then the conjugate momentum $\pi_Z, \pi_B$ are written as
\be
\left(
\begin{array}{c}
\pi_Z \\ \pi_B
\end{array} \right)
=\mbox{R}^{-1} \mbox{Diag}(\Psi_+,\Psi_-)
\left(
\begin{array}{c}
\Pi_+ \\ \Pi_-
\end{array} \right).
\ee
The boundary action is now written only by boundary values ($\hat{Z}_1, \hat{B}$) and conjugate momentum of master field $\Pi_\pm$. The two point function for $h^x_t$ and $h^x_z$ is related to the gauge invariant variable $Z_1$ as
\be
\mathcal{G}_{xtxt}= \frac{\delta^2 S_{bd}}{\delta \hat{h}^x_t \delta \hat{h}^x_t} = \left( \frac{\delta \hat{Z}_1}{\delta h^x_t}\right)^2 \frac{\delta^2 S_{bd}}{\delta \hat{Z}_1 \delta \hat{Z}_1}=\tilde{\qn}^2 \frac{\delta^2 S_{bd}}{\delta \hat{Z}_1 \delta \hat{Z}_1}, \quad \mathcal{G}_{xx}= \left(\frac{\delta \hat{B}}{\delta \hat{A}_x}\right)^2 \frac{\delta^2 S_{bd}}{\delta \hat{B} \delta \hat{B}}=\frac{1}{\mu^2}\frac{\delta^2 S_{bd}}{\delta \hat{B} \delta \hat{B}}
\ee
therefore the correlation functions for each components are
\ba
\mathcal{G}_{xt,xt}&=& \frac{l^3}{32 G_5^2 b^4}\tilde{\qn}^2 \frac{C_- \hat{\Pi}_+-C_+\hat{\Pi}_-}{C_+-C_-} ,\quad
\mathcal{G}_{xz,xz} =\frac{\wn^2}{\qn^2} \mathcal{G}_{xt,xt}\no
\mathcal{G}_{xt,x}&=& \mathcal{G}_{x,xt} = \tilde{\qn}^2 \frac{l^2 \sqrt{6a}}{32 G_5 e b^3}  \frac{\hat{\Pi}_+-\hat{\Pi}_-}{C_+-C_-} \no
\mathcal{G}_{x,x}&=& \frac{l}{4 e^2 b^2} \frac{C_+ \hat{\Pi}_+ - C_- \hat{\Pi}_-}{C_+-C_-} = \frac{l}{4 e^2 b^2}\frac{1}{2}\left(\frac{\hat{\Pi}_+ - \hat{\Pi}_-}{\gamma} + \hat{\Pi}_+ +\hat{\Pi}_-\right).
\ea
Note that when spatial momentum $\qn$ or density "a" vanishes two point function $\mathcal{G}_{xt,x}$ vanishes. It means that the holographic operator mixing between $Z_1$ and $A_x$ comes from the density effects. By following the standard recipe described in section 2, $\Pi_\pm$, the conjugate momentum of master fields, are computed as the ratio of two connection coefficients
\be \label{specexp1}
\Psi_\pm = \calA_\pm \bigg(1+\cdots \bigg) +\calB_\pm u \bigg(1+\cdots \bigg) =  \hat{\Psi}_{\pm} \left[1 + \cdots + \hat{\Pi}_{\pm}u \cdots \right]
\ee
By comparing eq. (\ref{specexp1}) with eq. (\ref{specexp1}) we get the conjugate momentum of the master fields as a ratio of connection coefficient of near boundary solutions of them,
\be
\Im~\frac{\calB_\pm}{\calA_\pm}  =\Im~\hat{\Pi}_\pm
\ee
By imposing infalling IR boundary condition for $\Psi_\pm$, the spectral functions are computed.
\be
\chi_{xtxt} = 2~\Im ~ \mathcal{G}_{xtxt},\quad \chi_{xzxz} =\frac{\wn^2}{\qn^2} \chi_{xtxt} ,\quad \chi_{x,xt} = \Im ~ \mathcal{G}_{xxt},\quad \chi_{xx} =2~\Im ~ \mathcal{G}_{xx}.
\ee
The spectral function is plotted in terms of $\wn, \qn$. The figure \ref{vsqnzero} shows that the imaginary part of the $G_{xx}$ divided by w, which is  AC conductivity of thermalized plasma ( with normalization constant, $\frac{1}{2}\frac{l}{g_5^2} 2\pi T$). The peak position becomes larger as the charge  increases.  The strength of that peak  also increases, when charge grows. In ref.\cite{Teaney1}, they calculate only zero density case which is  denoted  by  the thick line in fig. \ref{vsqnzero}. The right figure  in fig. \ref{vsqnzero} is the density dependence of DC conductivity. From the spectral function, DC conductivity can be computed by taking zero frequency limit, $\sigma_E = \lim_{w \rightarrow 0}\frac{\chi_{xx}(w,k=0)}{w}$. As density increases, it decreases and in  sufficiently large density regime, DC conductivity is  negligible. This is rather surprizing since Drude formula in Maxwell theory says the conductivity is proportional to the density of the charge carrier. It seems that interaction between the charge carriers dominates the abundancy effect.
Such  drastic reduction of the DC conductivity can be an another explanation of the jet quenching
phenomena which are different from the explanation in ref. \cite{sinzahed,herzog}.
In highly dense system, the strongly interacting plasma can not carry charge over long distance because of density effect. If this is the relevant mechanism, raising the temperature supresses the Jet quecnching in LHC since it reduces $\bar\mu \sim \mu/T$.

In fig. \ref{qvarchix}, we plot the spectral function $\chi_{xx}$ in terms of spatial momentum with fixed frequency. The left one shows thick, dashed, thin line corresponds to $\wn$ = 0.1, 0.3, 0.5 with $\bar{\mu}$=1 and the right of fig. \ref{qvarchix} also shows thick, dashed, thin line corresponds to $\bar{\mu}$=1,2,3 with $\wn$=0.1. These results can be interpreted as a inverse thermal screening length of the super Yang-Mills plasma. It is interesting that for the tensor and vector mode the peak position is different. For the $\wn=0$, $\chi_{xx}$ is zero so screening mass is zero. But for the finite $\wn$ there is the peak and the position is a function of both $\wn$ and $\bar{\mu}$. Because the diffusive nature affects the interactions inside the medium, the screening effect are more complicated.

\begin{figure}
\begin{center}
    \includegraphics[angle=0, width=0.48 \textwidth]{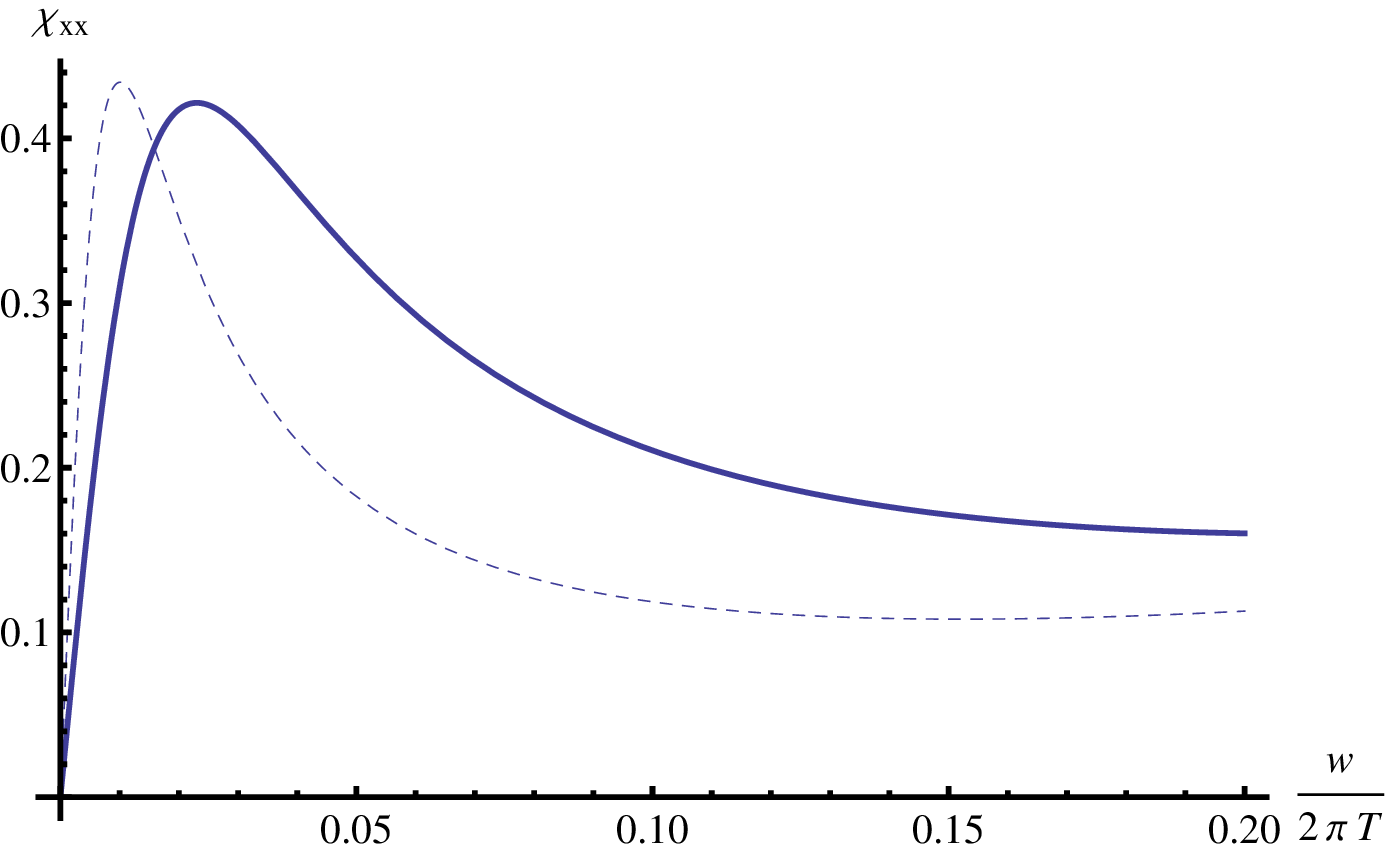}
    \includegraphics[angle=0, width=0.48 \textwidth]{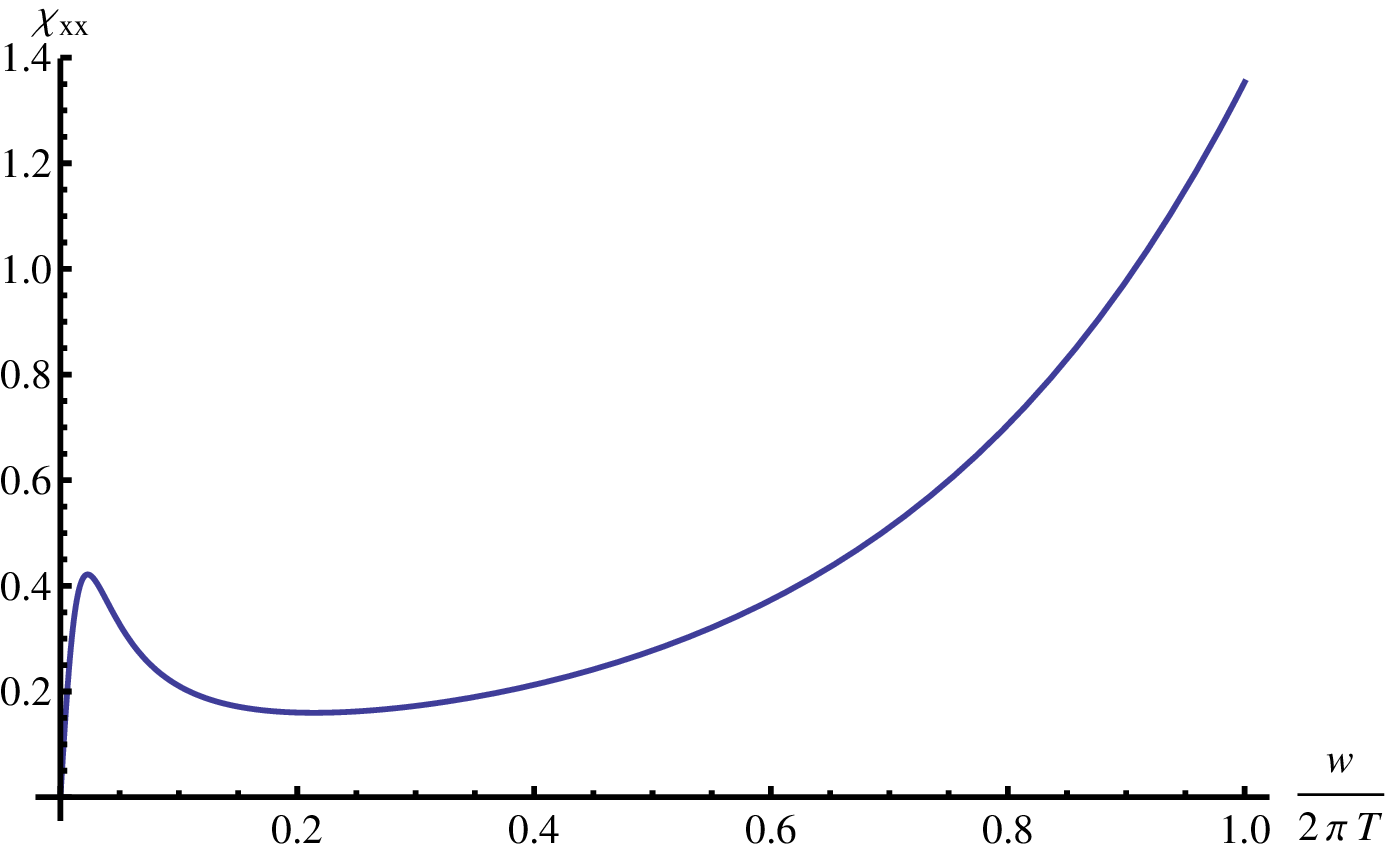}
  \caption{$\Delta\chi_{xx}$ for various density,   one with a=0.5, $\qn$=0.3 (thick), a=0.5 $\qn$=0.2(dashed). The normalization unit is $(2 \pi T)^2 \frac{l}{2e^2}$. Right : $\chi_{xx}$(a=0.5, $\qn$=0.3) plotted in range $\wn \in$ [0,1].}\label{xxhydropole}
\end{center}
\end{figure}
The hydrodynamic pole in $G_{xx}$ is appeared in fig \ref{xxhydropole} at
\be
\wn \sim -i \frac{1-a/2}{2(1+a)} \qn^2
\ee
The left figure shows that the hydrodynamic pole position is shifted from 0.0225 ($\qn$=0.3) to 0.01 ($\qn$=0.2). This comes from the density effect, when $\mu$ goes to zero the hydrodynamic pole in $G_{xx}$ disappears, see appendix \ref{smallwsp}. This is the operator mixing result. The diffusion pole is only appeared in $G_{xt,xt}$ or $G_{xz,xz}$ not $G_{x,x}$. The right figure shows that $\chi_{xx}$ reaches very rapidly to the zero temperature spectral function.

\begin{figure}
\begin{center}
    \includegraphics[angle=0, width=0.48 \textwidth]{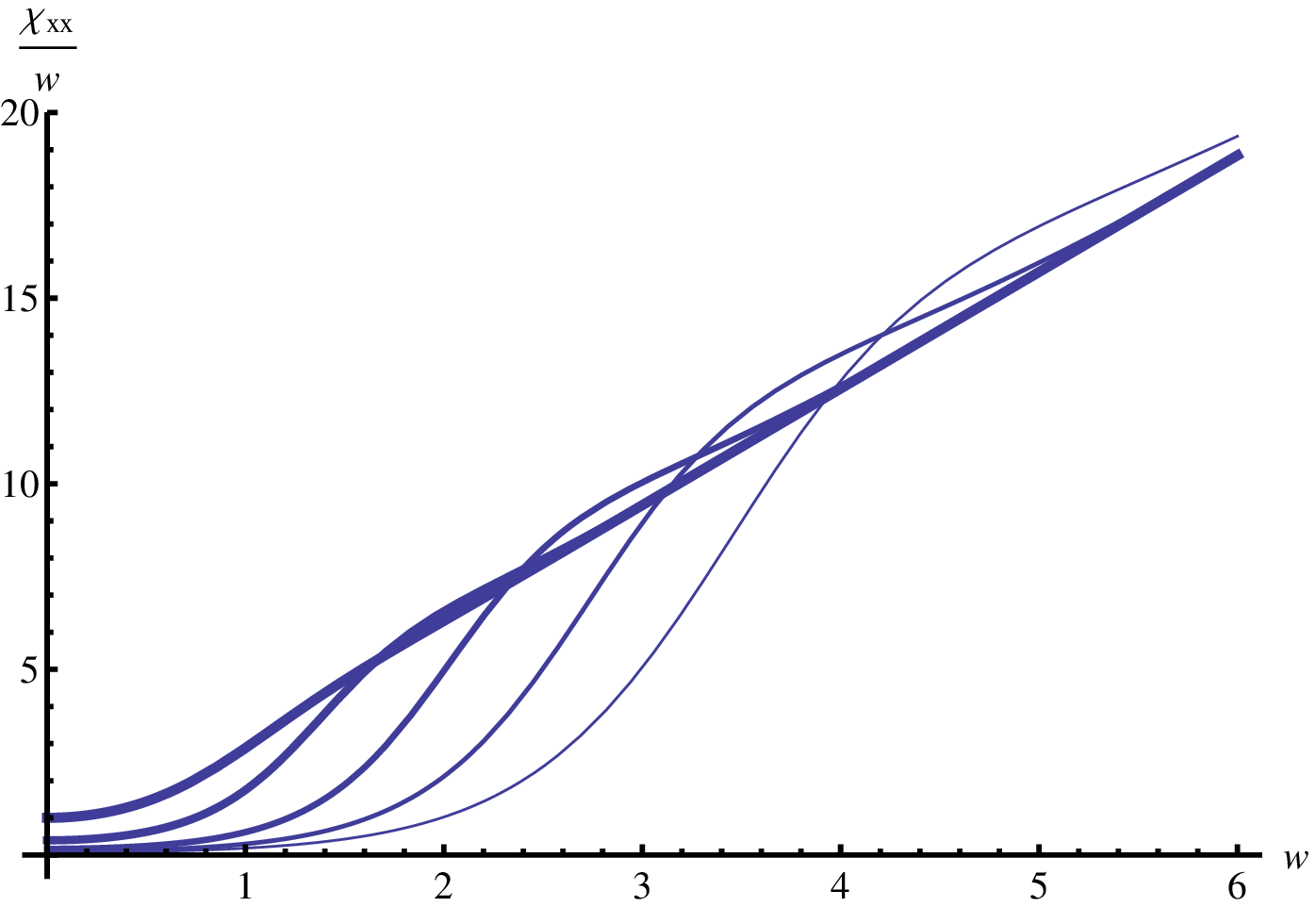}
    \includegraphics[angle=0, width=0.48 \textwidth]{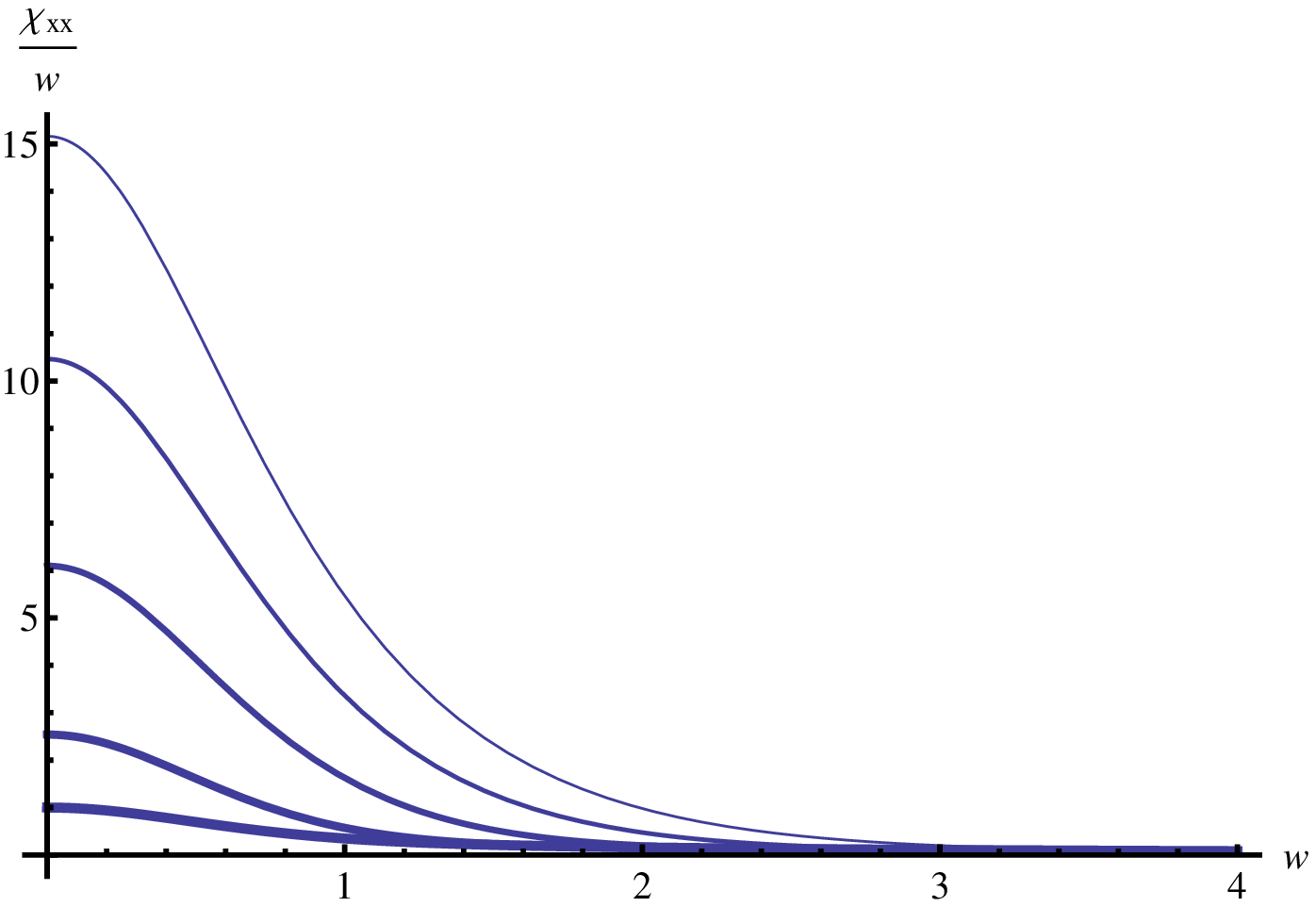}
  \caption{Real part of AC conductivity of SYM plasma, $\chi_{xx}(\wn,\qn=0)/\wn$ and the normalization unit is $(2 \pi T)^2 \frac{l}{2g_5^2}$. Each line shows the result when $\bar{\mu}$ = 0(thick), 1, 2, 3, 4(thin). Left : AC conductivity and Right : AC registivity.} \label{acconduct}
\end{center}
\end{figure}

Fig. \ref{acconduct} shows real part of ac conductivity, $\mbox{Re} \sigma(\wn)=\frac{\chi_{xx}(\wn,\qn=0)}{i \wn}$ with the normalization unit $(2 \pi T)^2 \frac{l}{2g_5^2}$. By definition, $\mbox{Im} \sigma = \frac{\mbox{Re} G_{xx}}{w}, \mbox{Re} \sigma = \frac{\mbox{Im} G_{xx}}{w}$. The ac registivity is defined as $\rho(\wn) = 1/\sigma(\wn)$. For large $\wn$, the system has zero registivity, it means that at any density charge carrying is almost perfect in high frequency.

\begin{figure}
\begin{center}
    \includegraphics[angle=0, width=0.45 \textwidth]{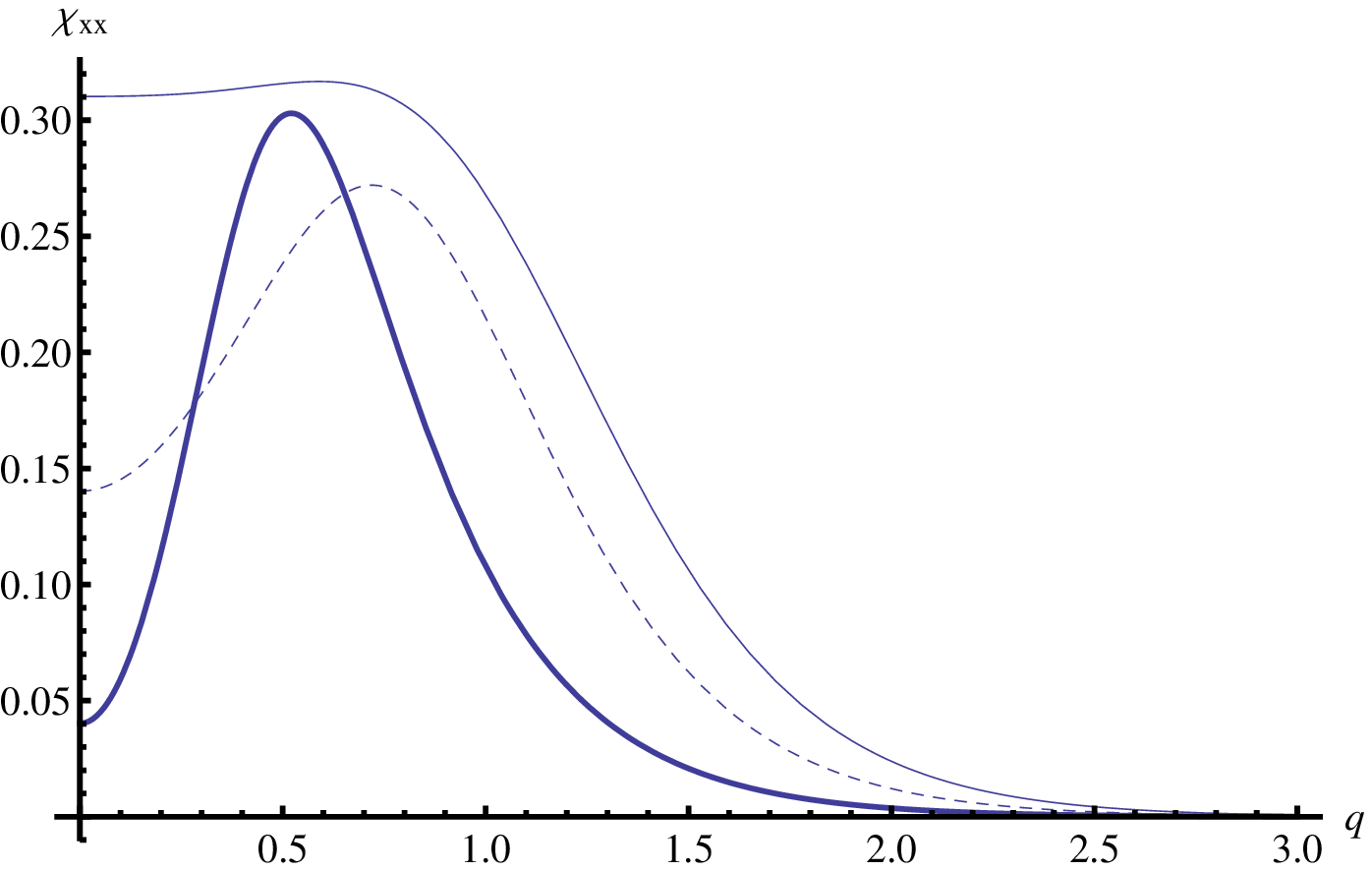}
    \includegraphics[angle=0, width=0.45 \textwidth]{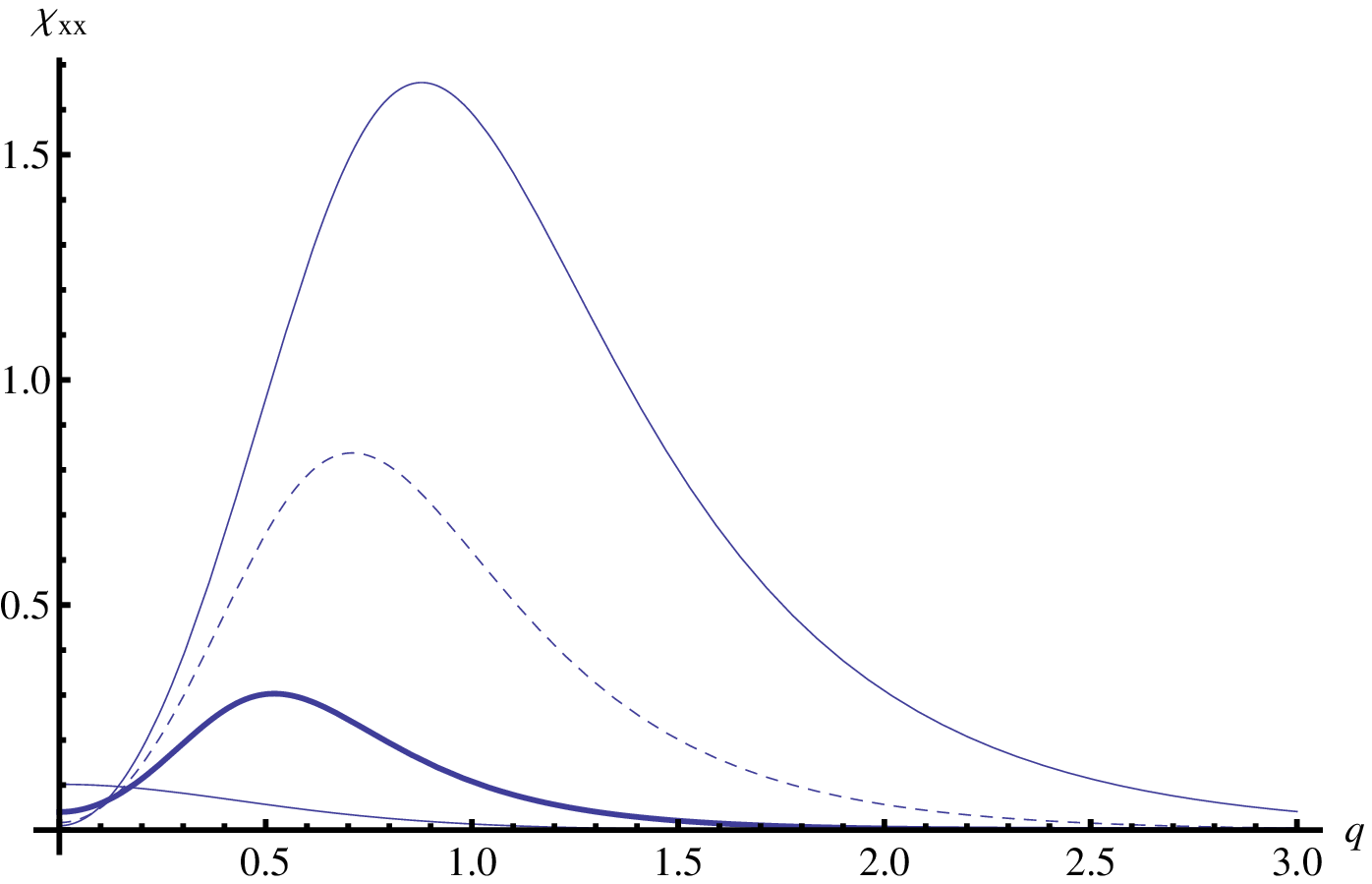}
  \caption{The normalized thermal spectral function $\chi_{xx}(\qn)$. Left: with $\bar{\mu}$=1 and varying $\wn$=0.1(thick), 0.3(dashed), 0.5(thin), Right : $\wn$=0.1 and varying $\bar{\mu}$=0(lowest thin line), 1(thick), 2(dashed), 3(solid),  with normalization unit $(2 \pi T)^2 \frac{l}{2g_5^2}$. }\label{qvarchix}
\end{center}
\end{figure}

\begin{figure}
\begin{center}
    \includegraphics[angle=0, width=0.48 \textwidth]{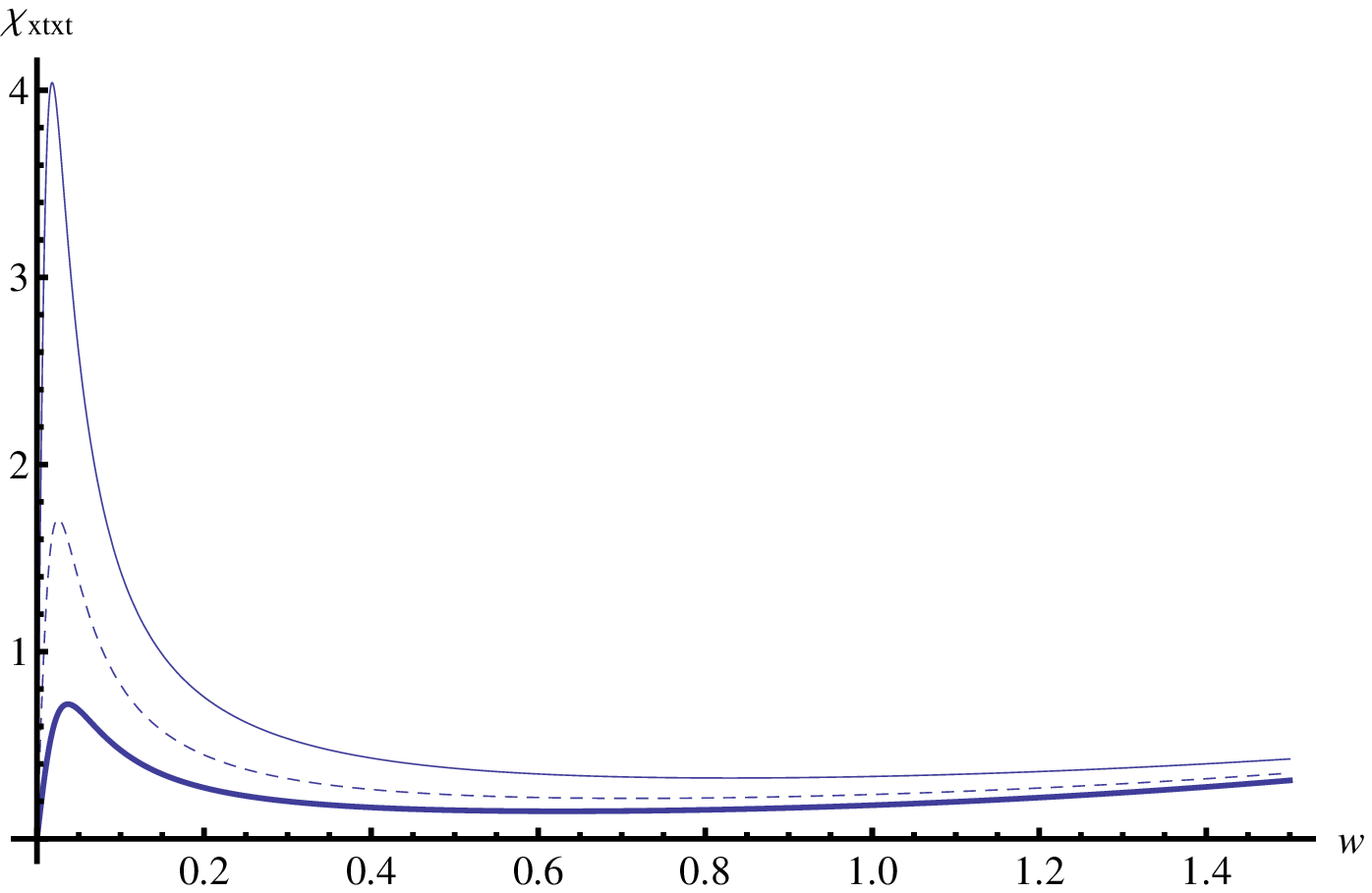}
    \includegraphics[angle=0, width=0.48 \textwidth]{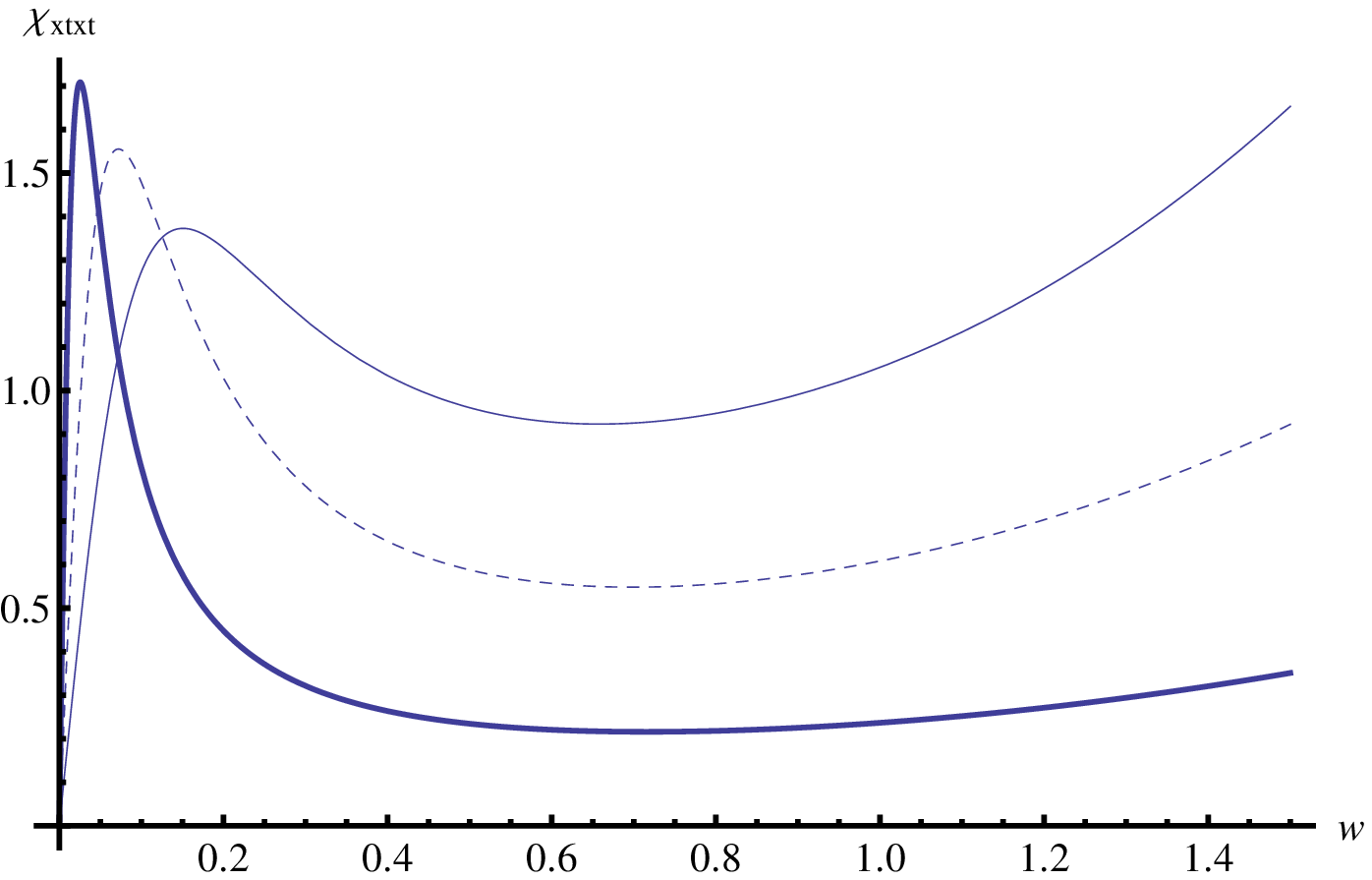}
  \caption{The normalized thermal spectral function $\chi_{xtxt}$. Left: with $\qn$=0.3 and varying $\bar{\mu}$=0.5(thick), 1(dashed), 1.5(solid), Right: with $\bar{\mu}$=1 and varying $\qn$=0.3(thick), 0.5(dashed), 0.7(solid) with normalization unit $(2 \pi T)^4 \frac{l^3}{16G_5^2}$. } \label{spxtxtvar}
\end{center}
\end{figure}

The (xt,xt) component of spectral function $\chi_{xtxt}$ has the diffusion pole at $\wn = \qn^2/2$ \cite{Kovtun:2005ev}. The dispersion relation for diffusive channel $w=D k^2/2$ gives us the diffusion constant $D=1/2\pi T$ from the hydrodynamic analysis. The left figure in fig. \ref{spxtxtvar} is with $\qn$=0.3 for various  values of $\bar{\mu}$:  0.5(thick), 1(dashed), 1.5(thin). When the chemical potential grows , the strength of peak also grows but   the position itself does not. The reason of this increase comes from the factor $\frac{1}{(1-a/2)^4}$ in front of the Greens function. When $\bar{\mu}$   increases the parameter $a$ goes up so the overall factor $(1-a/2)^{-4}$  increases rapidly.   When the system reaches the extremal limit, $a=2$, that factor diverges and our analysis is broken down. It should be computed separately for the zero temperature or for the extremal RN spectral function from finite temperature or non extremal RN AdS blackhole.

In the right figure  of the fig. \ref{spxtxtvar}, the peak position is shifted when $\qn$ is moved. Again, the position is at $\wn = \qn^2/2$.

\section{Photo-emission rate}
\begin{figure}
\begin{center}
    \includegraphics[angle=0, width=0.48 \textwidth]{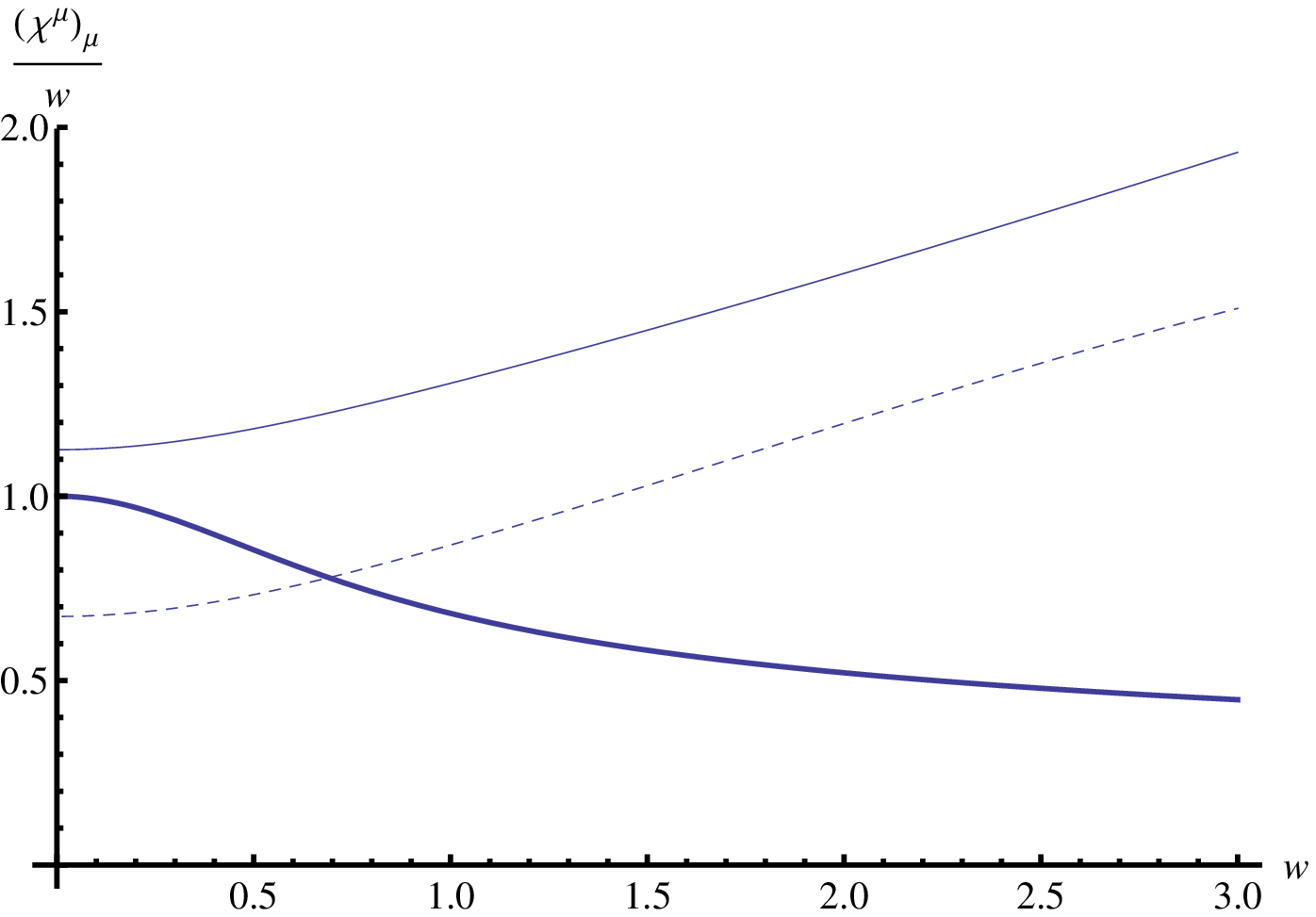}
    \includegraphics[angle=0, width=0.48 \textwidth]{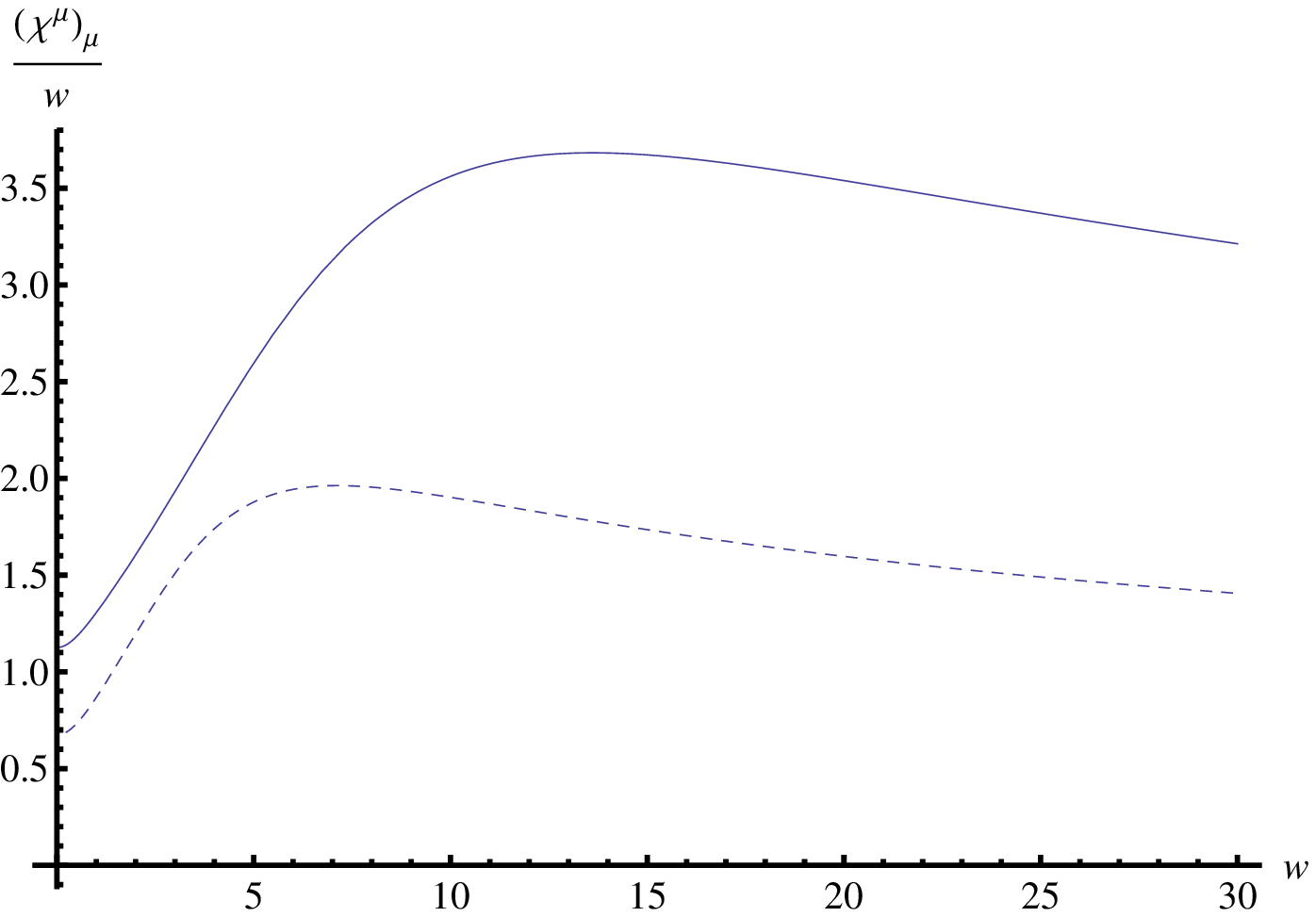}
  \caption{The normalized trace of thermal spectral function $\chi^\mu_\mu/\wn$ with light like momenta. Left : varying $\bar{\mu}$=0(thick), 5(dashed), 10(thin). Right : $\bar{\mu}$= 5(dashed), 10(thin) and the plotting range is from zero to 30 in $\wn$. }\label{chixweqq}
\end{center}
\end{figure}

In the heavy ion collision, the emitted photons are a good measure to see the medium effect. The photo emission rate of SYM plasma was  calculated holographically for AdS Schwarzschild \cite{CaronHuot:2006te}, for D4/D8/$\bar{D8}$ with finite chemical potential \cite{Parnachev:2006ev} and for D3/D7 with finite baryon density \cite{Mas:2008jz}. We will focus on the photo-emission rate for our gauge theory dual to the RN AdS background here. Let $\Gamma_\gamma$ be the number of photons emitted per unit volume.  To  leading order in electromagnetic coupling e,
\ba
d\Gamma_\gamma &=& \frac{d^3 k}{(2\pi)^3} \frac{e^2}{2 |\vec{k}|} \eta^{\mu\nu} C^{<}_{\mu\nu}(K)|_{w=k} \no
C^{<}_{\mu\nu}(K) &=& n_B(w)\chi_{\mu\nu}(K)
\ea
where $C^{<}_{\mu\nu}(K)$ is the Fourier transformed Wightman function which is related to the spectral function multiplied by Bose-Einstein distribution function $n_B(w)$. Convert the differential photo-emission rate into the emission rate per unit volume as a function of $\omega$,
\be
\frac{d \Gamma_\gamma}{dk} = \frac{\alpha_{EM}}{\pi}~k~ \eta^{\mu\nu}C^{<}_{\mu\nu}(K)|_{\omega=k}.
\ee
In order to calculate photo-emission rate we need to know the longitudinal part of the spectral function. But for the case of light like momentum $\omega=k$, the trace of spectral function $\chi^\mu_\mu = \eta^{\mu\nu}\chi_{\mu\nu}$ is obtained only by $\Pi^T$. From the appendix \ref{Appd1},
\be
\eta^{\mu\nu}C^{ret}_{\mu\nu} = (d-2)\Pi^T+\Pi^L = 2\Pi^T+\Pi^L
\ee
where d is the dimension of the boundary field theory. For the light like momentum, the longitudinal correlator should be vanished because the projection operator diverges. The trace of spectral function is only given by $\Pi^T(w,|\vec{k}|=w)$. The frequency dependent spectral measure $\eta^{\mu\nu}\chi_{\mu\nu}/\wn$ for the light like momenta is in fig. \ref{chixweqq}. The left one in fig. \ref{chixweqq} has three lines $\bar{\mu}$ = 0(thick), 5(dashed), 10(thin).

In figure 7, the photo-emission rate is  compared with  \cite{CaronHuot:2006te}, the peak position is $\wn_{max} =1.48479/(2\pi)=0.2363$  the maximum value is 0.01567 with unit $\alpha_{EM}(N_c^2-1)T^3$. The left figure shows the photo emission rate for  $\bar{\mu}$=0(thick), 1(solid), 5(dashed), 10(thin) and the right shows the maximum value $\frac{d\Gamma_\gamma}{dk}(w=0.2363)$ as a function of $\bar{\mu}$. Notice that the maximum value of the photo emission rate  decreases until $\wn_c$ = 2.014 which  is the turning point. After passing $\bar{\mu}_c$ it   increases. It means the thermal photon production rate is suppressed in low chemical potential regime but in high density regime it is enhanced. It may reflects the fact that at high enough density thermal screening is enhanced. As we have seen the suppression of conductivity and increase of thermal screening mass in fig. \ref{vsqnzero}, \ref{qvarchix}.
\begin{figure}
\begin{center}
    \includegraphics[angle=0, width=0.48 \textwidth]{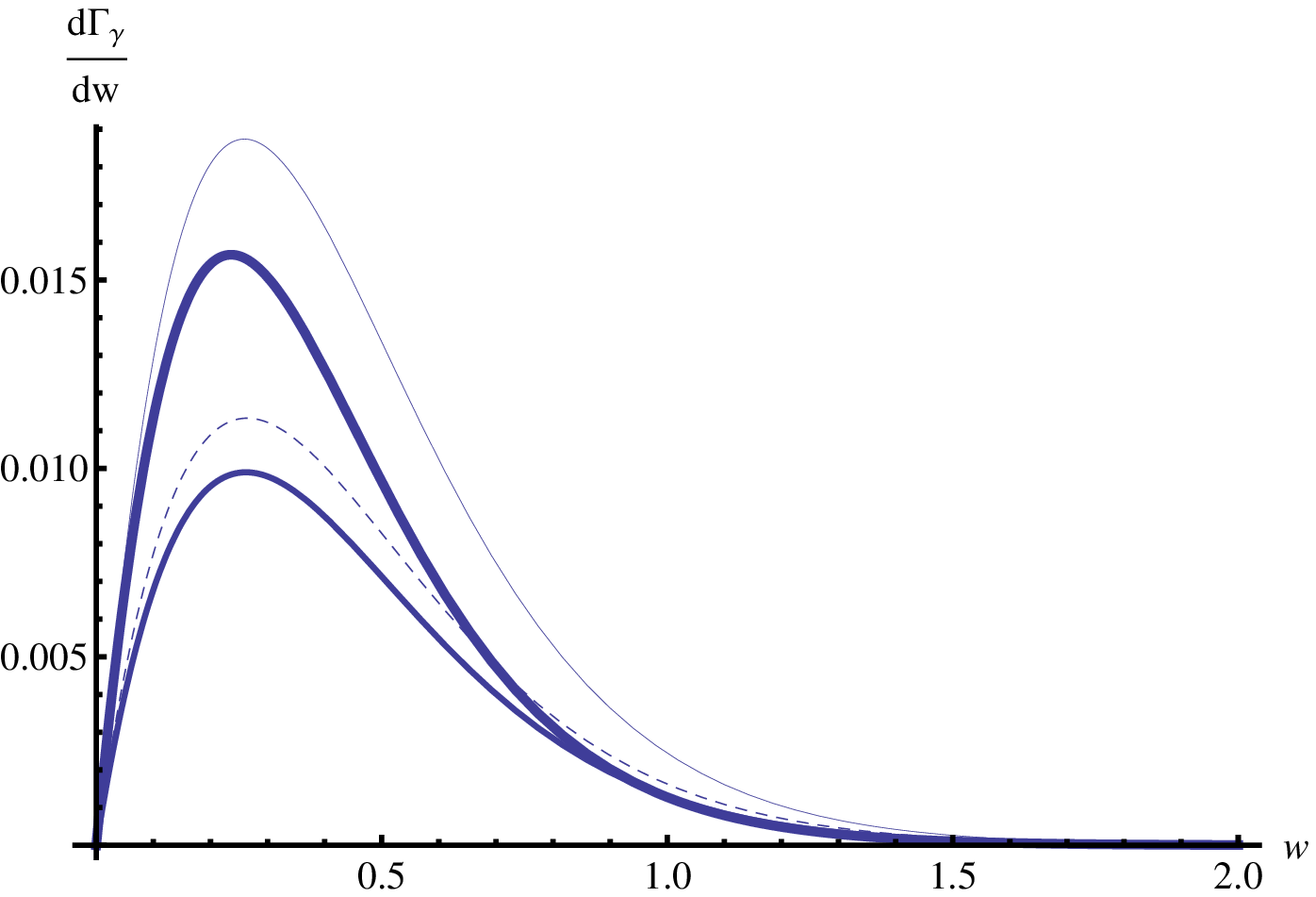}
    \includegraphics[angle=0, width=0.48 \textwidth]{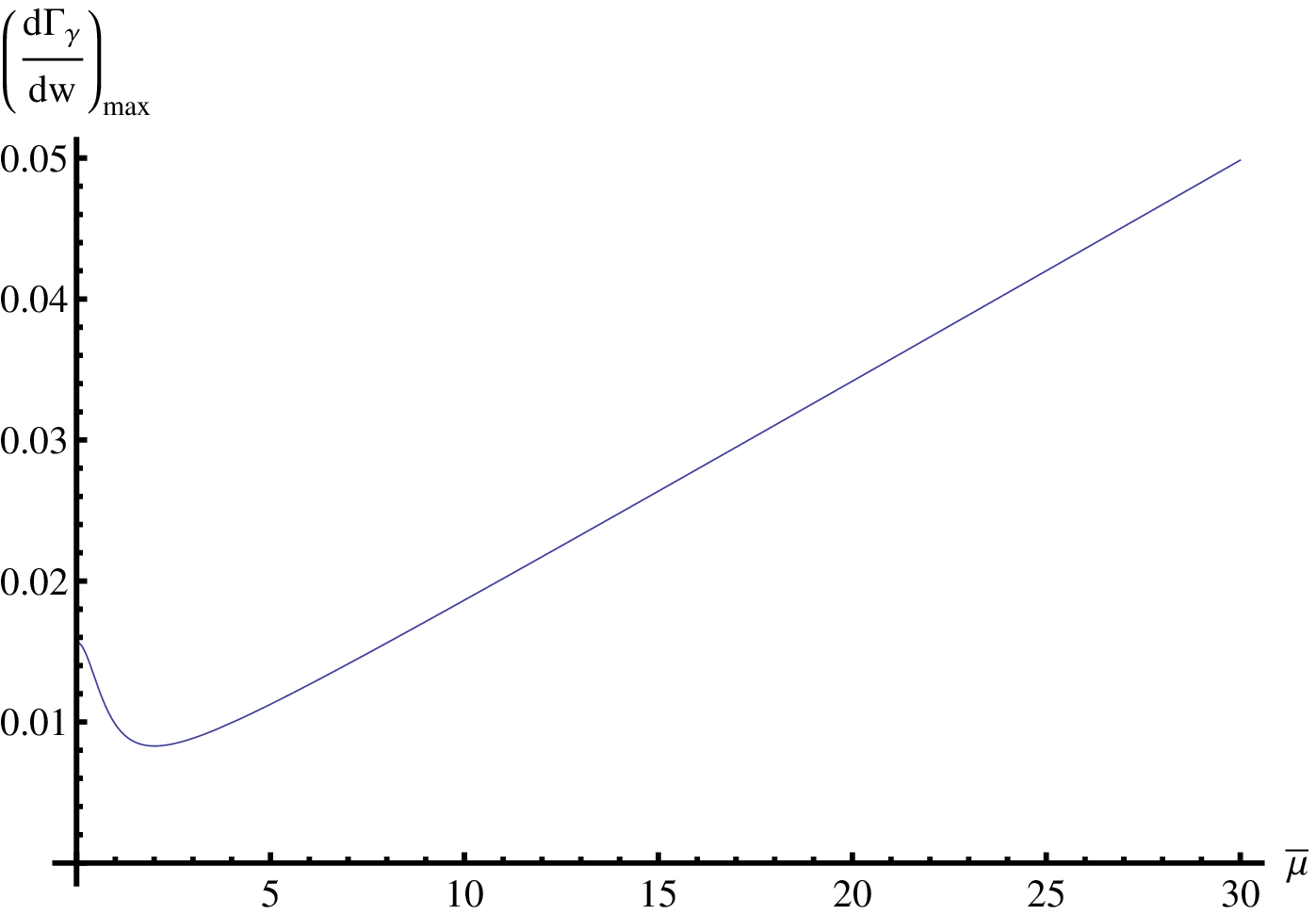}
  \caption{The photo emission rate of SYM-EM plasma with normalization unit $\alpha_{EM}(N_c^2-1)T^3$. $d\Gamma_\gamma/dk$ with light like momenta. Left: varying $\bar{\mu}$=0(thick), 1(solid), 5(dashed), 10(thin), Right : The maximum value of the photo emission rate as a function of chemical potential $\bar{\mu}$.}\label{photoemission}
\end{center}
\end{figure}
When the U(1) chemical potential is very large, $a\sim 2$, the maximum value suddenly increase. Until a=1, the maximum value is slowly decreasing function of a, but after passing a=1 it increases stiffly. But this sudden increase may be artificial, because if we express   as the chemical potential this stiffness is not there.
\begin{figure}
\begin{center}
    \includegraphics[angle=0, width=0.48 \textwidth]{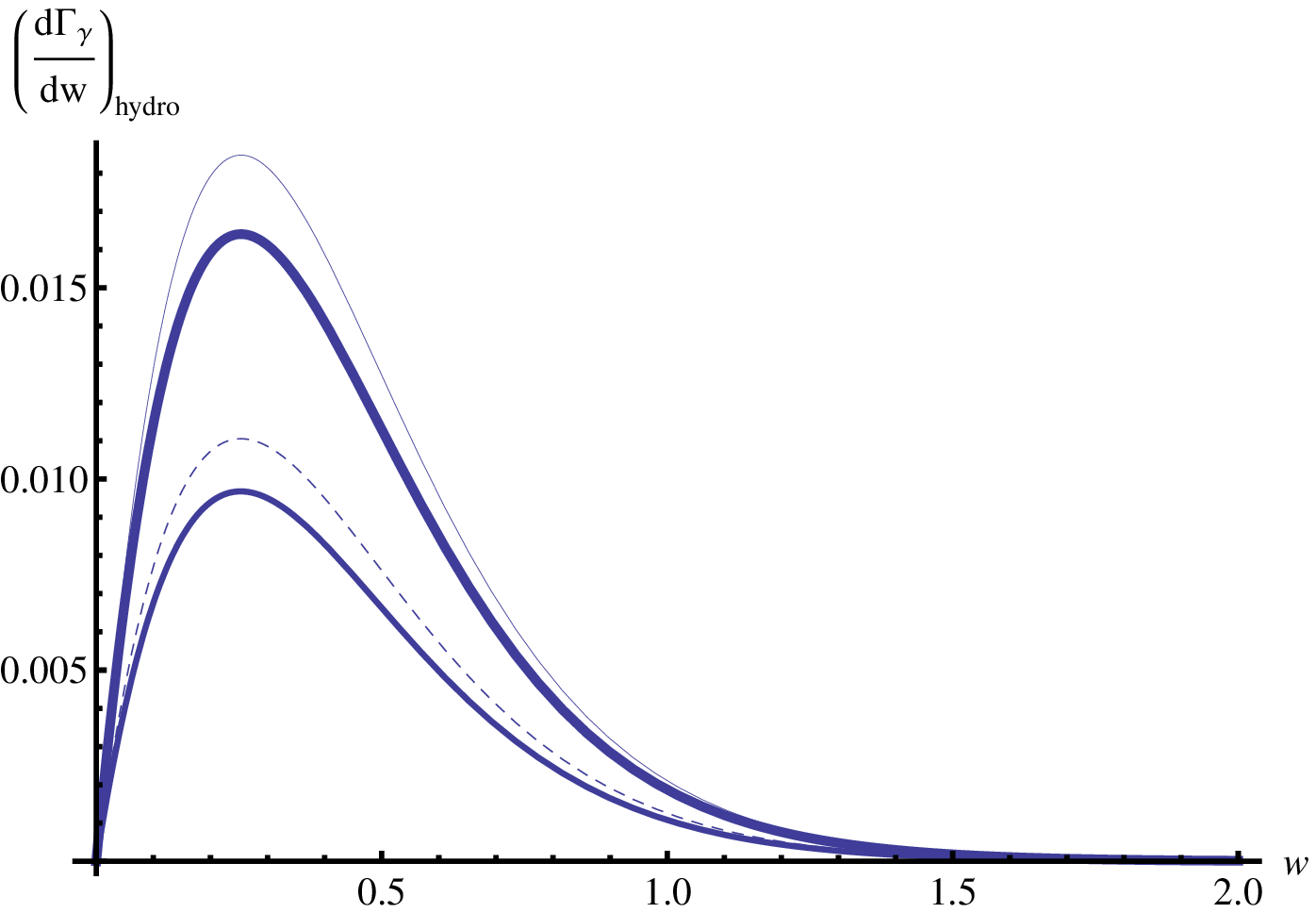}
    \includegraphics[angle=0, width=0.48 \textwidth]{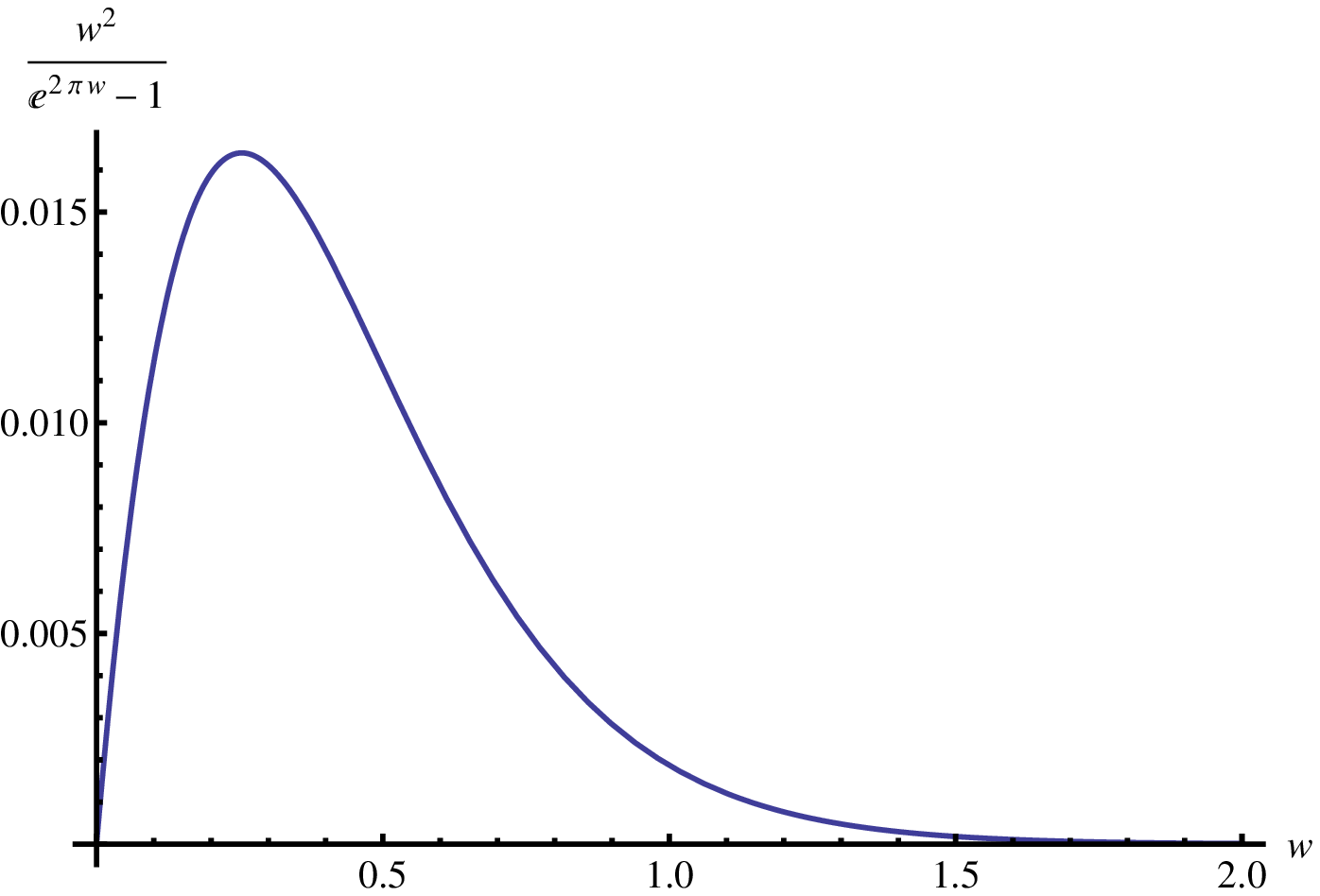}
  \caption{The photo emission rate of hydrodynamic approximated spectral function with normalization unit $(2\pi T)^2\frac{l}{2g_5^2}$. $d\Gamma_\gamma/dk$ with light like momenta. Left: varying $\bar{\mu}$=0(thick), 1(solid), 5(dashed), 10(thin), Right : $\wn^2/(\e^{2\pi \wn}-1)$ is plotted as a function of $\wn$. }\label{analytphoemden}
\end{center}
\end{figure}

Note that the spectral measure has maximum at the point due to the hydrodynamic pole of $G_{xx}$.
\be
G_{xx} \sim \frac{1}{w+i Dk^2}, \quad \mbox{where} \quad D = \frac{b}{2(1+a)}
\ee
A simple analytic function, see appendix \ref{smallwsp}, which has only above hydrodynamic pole structure present the very essence of photoemission rate of dense supersymmetric Yang-Mills plasma.
\ba \label{hydrospectraldens}
\chi^{hydro}_{xx} &=& \frac{2l}{8e^2b^2} \bigg[\frac{3a}{1+a}\frac{Dk^2w}{D^2k^4+w^2}+\frac{2(1-a/2)^2}{(1+a)^2}b w \bigg] \no
&=&\frac{2l}{8e^2b^2} \bigg[\frac{3a}{1+a}\frac{D_a \tilde{\qn}^2 \tilde{\wn}}{D_a^2 \tilde{\qn}^4+\tilde{\wn}^2}+\frac{2(1-a/2)^3}{(1+a)^2}\tilde{\wn} \bigg], \quad D_a = \frac{1}{2(1+a)}
\ea
Fig. \ref{analytphoemden} shows that the photo emission rate of hydrodynamic approximated solution. The left one is almost same as full numerical solution but tails in large $\wn$ are different and the right one we plot $\wn^2/(\e^{2\pi \wn}-1)$ as a function of $\wn$. The origin of peak in photo emission rate comes from the statistical factor and the change of height is density effect. This density effect is almost described by a single function eq. (\ref{hydrospectraldens}) \footnote{But actually at large $\wn$, there are deviations between fig.\ref{photoemission} and \ref{analytphoemden}. Please note that it is an approximated solution.  }.

\section{Conclusion}
We have solved the equation of motion for linerized gravitational and electromagnetic perturbations in RN AdS background to get the holographic spectral function. Due to the density effect, the gravitational and electromagnetic perturbations are coupled with each other so it is not easy to solve. By introducing master variable, however, we can decouple these modes which makes the problem simpler. The problem  might  be handled without decoupling along the method discussed in  \cite{Kaminski:2009dh}.

The density effects of thermal spectral function have some interesting features. The boundary theory of RN AdS is believed supersymmetric Yang-Mills theory with finite U(1) charge density. The original SYM  has no dimensionful parameter and does not admit any quantity like screening length, energy gap, diffusion constant etc. In the finite temperature and density state, however, the plasma have a scale given by those. The temperature introduce diffusion constant, conductivity and the density modify these quantities.

One of the interesting feature is the modification of the diffusive nature of thermal plasma. As we have seen the fig. \ref{vsqnzero} the DC conductivity is decreasing in large $\bar{\mu}$. It is quite natural because in  dense medium, particles collide very frequently so   charge carrying process should be suppressed.

The photo emission rate is very important tool to probe the effects of thermal medium. Because photon does not interacted with other particles via strong interaction, it carries informations of the early stage of collision. Holographic photo emission rate is greatly enhanced when $\bar{\mu}$ is very large. See figure \ref{photoemission}.
Note that  information of spectral function over only  small window of $\wn$ is sufficient to describe the photo emission rate, because almost all of the contribution for photoemission rate comes from the statistical factor $1/(\e^{w/T}-1)$.

It would be interesting to see whether  the U(1) axial anomaly \cite{Sahoo:2009yq,Yee:2009vw,Matsuo:2009xn} can affect the  photoemission rate. We will report this issue in  the near future.\\

\noindent{\bf \large Acknowledgments}
%\noindent
This work was supported  by the WCU project of Korean Ministry of Education, Science and Technology (R33-2008-000-10087- 0). This work is also  supported  by the National Research Foundation of Korea(NRF) grant funded by the Korea government(MEST) (No.2005-0049409). Also we give thanks to Yang Zhou, Alberto and Antonio to inform to correct some typos.

\appendix

\section{The index structure of correlation functions} \label{Appd1}
The correlation functions of various operators has Lorentz index structure, and that has to satisfy some constraints, ie. CPT invariance, Ward identity \cite{Kovtun:2005ev}. We will briefly review the index structure of our correlation functions for later convenience.

The definition of retarded correlation function of conserved current and energy momentum tensor are
\ba
C_{\mu\nu} (x-y) &=& -i \theta(x^0-y^0)\left<[J_\mu(x),J_\nu(y)]\right> \no
G_{\mu\nu\alpha\beta} (x-y) &=&-i \theta(x^0-y^0)\left<[T_{\mu\nu}(x),T_{\alpha\beta}(y)]\right>,
\ea
the state are translational invariant so we can Fourier transform these correlation functions into momentum space.
In the equilibrium, CPT invariance told us that
\be
C_{\mu\nu} = C_{\nu\mu}, \quad G_{\mu\nu\alpha\beta}= G_{\alpha\beta\mu\nu}
\ee
in addition correlation functions of energy momentum tensor has the property inherited from the symmetry of energy momentum tensor
\be
G_{\mu\nu\alpha\beta} = G_{\nu\mu\alpha\beta}= G_{\mu\nu\beta\alpha}.
\ee
The Ward identity,
\be
k^\mu C_{\mu\nu} = 0 = k^\mu G_{\mu\nu\alpha\beta}
\ee
and if the theory has scale invariance $T^\mu_\mu=0$,
\be
\eta^{\mu\nu} G_{\mu\nu\alpha\beta}=0.
\ee
From the Ward identity, the correlation functions are projected onto transverse spacetime of $k^\mu$
\ba
C_{\mu\nu} &=& P_{\mu\nu} \Pi(K^2) \no
G_{\mu\nu\alpha\beta} &=& P_{\mu\nu} P_{\alpha\beta} G_B(K^2) + H_{\mu\nu\alpha\beta} G_S(K^2)
\ea
where
\ba
P_{\mu\nu} &=& \eta_{\mu\nu} - \frac{k_\mu k_\nu}{K^2} \no
H_{\mu\nu\alpha\beta} &=& \frac{1}{2}(P_{\mu\alpha}P_{\nu\beta}+P_{\mu\beta}P_{\nu\alpha})-\frac{1}{D-1}P_{\mu\nu}P_{\alpha\beta}
\ea

The field theory propagator is classified by boundary SO(2) rotation symmetry \cite{Kovtun:2005ev}. The propagator is decomposed as scalar, vector, tensor parts according to their transformation properties under SO(2). We assume that the wave is going along z direction $K=(w,0,0,k)$ and the boundary coordinate is labeled by (t,x,y,z). The stress-energy tensor correlator $G_{\mu\nu\rho\sigma}$ is also decomposed into these categories. The conserved current, for our case R-current, is projected as transverse and longitudinal parts
\be
C_{\mu\nu} = P^T_{\mu\nu} \Pi^T +P^L_{\mu\nu} \Pi^L
\ee
where $P^T_{\mu\nu}, P^L_{\mu\nu}$ is transverse and longitudinal projector which are mutually orthogonal
\be
P^T_{ij} = \delta_{ij}-\frac{k_i k_j}{\vec{k}^2}, \quad P^T_{\mu 0}=P^T_{0 \mu}=0 ,\quad P^L_{\mu\nu}=P_{\mu\nu}-P^T_{\mu\nu}.
\ee
Each component of the current-current correlator is
\ba
C_{xx}(K)&=&C_{yy}(K)=\Pi^T(K) \no
C_{tt}(K)&=&\frac{k^2}{w^2-k^2}\Pi^L,\quad C_{tz}(K)=\frac{-wk}{w^2-k^2}\Pi^L,\quad C_{zz}(K)=\frac{w^2}{w^2-k^2}\Pi^L.
\ea
For the stress-energy correlation function, the classification is slightly complicated.
\ba
G_{\mu\nu\alpha\beta}(K) &=& \left(P^T_{\mu\nu} P^T_{\alpha\beta}+\frac{1}{2}(P^T_{\mu\nu}P^L_{\alpha\beta}+P^L_{\mu\nu}P^T_{\alpha\beta}) \right)C_T \no
&&+\left(P^T_{\mu\nu} P^T_{\alpha\beta}+\frac{1}{2}(P^T_{\mu\nu}P^L_{\alpha\beta}+P^L_{\mu\nu}P^T_{\alpha\beta}) \right)C_L \no
&& + S_{\mu\nu\alpha\beta}G_1 + Q_{\mu\nu\alpha\beta}G_2 + L_{\mu\nu\alpha\beta}G_3,
\ea
where
\ba
S_{\mu\nu\alpha\beta} &=& \frac{1}{2} \bigg(P^T_{\mu\alpha}P^L_{\nu\beta}+P^T_{\mu\alpha}P^L_{\nu\beta}+P^T_{\mu\beta}P^L_{\nu\alpha}+P^L_{\mu\beta}P^T_{\nu\alpha}\bigg) \no
Q_{\mu\nu\alpha\beta} &=& \frac{1}{D-1}\bigg((D-2)P^L_{\mu\nu}P^L_{\alpha\beta}+\frac{1}{D-2}P^T_{\mu\nu}P^T_{\alpha\beta}-(P^T_{\mu\nu}P^L_{\alpha\beta}+P^L_{\mu\nu}P^T_{\alpha\beta})\bigg) \no
L_{\mu\nu\alpha\beta} &\equiv& H_{\mu\nu\alpha\beta}-S_{\mu\nu\alpha\beta}-Q_{\mu\nu\alpha\beta} \no
\ea
The transverse component of $G_{\mu\nu\alpha\beta}$ are
\ba
G_{txtx}&=&\frac{1}{2}\frac{k^2}{w^2-k^2}G_1, \quad G_{txzx}=-\frac{1}{2}\frac{wk}{w^2-k^2}G_1, \no
G_{xzxz}&=&\frac{1}{2}\frac{w^2}{w^2-k^2}G_1, \quad G_{xyxy}=\frac{1}{2}G_3
\ea
and longitudinal component of G are
\ba
G_{tttt} &=& \frac{1}{3}\frac{k^4}{(w^2-k^2)^2}\bigg[2G_2+3C_L\bigg] \quad
G_{tttz} = -\frac{1}{3}\frac{wk^3}{(w^2-k^2)^2}\bigg[2G_2+3C_L\bigg] \no
G_{ttxx} &=& \frac{1}{6}\frac{k^2}{w^2-k^2}\bigg[2G_2-3C_L-3C_T\bigg].
\ea
Note that if the theory has the scale invariance, $C_L, C_T$ vanish.

\section{Large $\wn$ spectral funtions} \label{largewsp}
In this section, we will calculate the spectral function analytically in the low and large frequency limit \cite{Atmaja:2008mt, Teaney1}. By using WKB methods, we can get the large frequency spectral function. This procedure is easily described by an example of simple Schrodinger equation,
\be
{y}''(x)+V(x) y(x) =0
\ee
here using this ansatz $y(x) = \mathrm{e}^{i \phi(x)}$ we get the equation for $\phi(x)$
\be \label{WKBeomphi}
-(\phi')^2 + i \phi'' + V=0 .
\ee
Assume that $\phi''$ is subleading,
\be
{\phi}'= \pm \sqrt{V}, \quad |\phi''| \sim \frac{1}{2} \left|\frac{V'}{\sqrt{V}}\right| << |V|
\ee
so first order solution is
\be
\phi = \pm \int \sqrt{V} dx.
\ee
From eq. (\ref{WKBeomphi}) we will obtain second order solution by substitute the first order solution,
\ba
(\phi')^2 & \sim & V \pm \frac{i}{2} \frac{V'}{\sqrt{V}} \no
\phi' & \sim & \pm \sqrt{V} + \frac{i}{4} \frac{V'}{V} \no
\phi & =& \pm \int \sqrt{V} dx + \frac{i}{4}\ln V
\ea
so WKB solution is
\be
y(x) = \frac{1}{V^{1/4}} \bigg(\e^{i \int \sqrt{V}dx} +\e^{-i \int \sqrt{V}dx}\bigg).
\ee

\subsection{Tensor mode}
For the tensor mode, the equation of motion for $h^x_y$ is transformed into Schrodinger form by choosing  $h^x_y(u)= X(u) \psi(u)$ where X(u) = $\sqrt{u/f(u)} $.
\be
{\psi}'' + V(u) \psi =0, \quad V(u) = -\frac{3}{4u^2}+ \frac{1}{4}\frac{f'^2}{f^2} + \frac{f'}{2uf} - \frac{f''}{2f}+\frac{\left(1-\frac{a}{2}\right)^2}{uf^2}(\wn^2-\qn^2 f)
\ee
This equation has two singular point at u=0 and u=1. From the WKB analysis, we get the two linearly independent solution away from these singularities
\be
\psi_1 \sim \frac{1}{\sqrt{p(u)}} ~\mathrm{cos}(S(u)+\phi_1), \quad
\psi_2 \sim \frac{1}{\sqrt{p(u)}} ~\mathrm{sin}(S(u)+\phi_2)
\ee
where
\be
p(u) = \frac{1-\frac{a}{2}}{\sqrt{u} f} \sqrt{\wn^2-\qn^2 f},\quad S(u) = \int^u_0 p(z) dz
\ee

\noindent
{\underline{\textbf{Near the boundary}}}\\
Near u=0, the potential has the form
\be
{\psi}'' + \bigg(-\frac{3}{4u^2}+\frac{Q^2}{u}\bigg)\psi =0
\ee
where $Q^2 = \left(1-\frac{a}{2}\right)^2(\wn^2-\qn^2)$. The general solution is
\be
\psi = C_1 \sqrt{u} J_2 (\sqrt{4Q^2 u})+ C_2 \sqrt{u} Y_2 (\sqrt{4Q^2 u}).
\ee
Bessel functions have the following asymptotic forms for large x (x $>>|\alpha^2-1/4|$)
\be
J_\alpha(x) \sim \sqrt{\frac{2}{\pi x}} \mathrm{cos} \left(x-\frac{\alpha \pi}{2} - \frac{\pi}{4}\right), \quad Y_\alpha(x) \sim \sqrt{\frac{2}{\pi x}} \mathrm{sin} \left(x-\frac{\alpha \pi}{2} - \frac{\pi}{4}\right).
\ee
We are considering large $\wn$, so asymptotic forms of Bessel is valid for our case,
\ba
\sqrt{u} J_2 (\sqrt{4Q^2 u}) & \sim & \sqrt{u} \frac{1}{\sqrt{\pi}} \frac{1}{\sqrt{Q \sqrt{u}}} \mathrm{cos} \left(2 Q \sqrt{u}- \frac{5\pi}{4}\right) \no
\sqrt{u} Y_2 (\sqrt{4Q^2 u}) & \sim & \sqrt{u} \frac{1}{\sqrt{\pi}} \frac{1}{\sqrt{Q \sqrt{u}}} \mathrm{sin} \left(2 Q \sqrt{u}- \frac{5\pi}{4}\right)
\ea
and these are well matched with our WKB solutions p(u) $\sim Q/\sqrt{u}$, S(u) $\sim 2 Q \sqrt{u}$.
{\underline{\textbf{Near the horizon}}}\\
Near u=1, the equation of motion is
\be
\psi'' + \frac{1}{(1-u)^2}\frac{1+\wn^2}{4}
\ee
and the solution is
\be
\psi = C_3 (1-u)^{1/2(1-i \wn)} +C_4 (1-u)^{1/2(1+i \wn)}.
\ee
Near the horizon, infalling wave is only physically relevant and this boundary condition is to choose $C_4$ as zero. We know the two asymptotic solutions and it should be matched at some point,
\be
\frac{1}{\sqrt{\pi p(u)}} \bigg[\mathrm{cos}\left(S(u)-\frac{5}{4\pi}\right) + i ~ \mathrm{sin}\left(S(u)-\frac{5}{4\pi}\right)\bigg] = C (1-u)^{1/2-i \wn/2}
\ee
from this, we get the solution for original field $h^x_y = X \psi$
\be
i~ u J_2 (\sqrt{4Q^2 u}) +u Y_2 (\sqrt{4Q^2 u}) = C (1-u)^{-i \wn/2}.
\ee
This is the solution for large $\wn, \qn$ with infalling boundary condition at the black hole horizon. We did not fix the coefficient C yet, it is determined by the condition $h^x_y(u=0)=1$. Near the boundary, Bessel functions has series solutions and
\be
u J_2 (\sqrt{4Q^2 u}) \sim \frac{1}{2} Q^2 u + O(u^3), \quad
u Y_2 (\sqrt{4Q^2 u}) \sim -\frac{1}{\pi Q^2}  + O(u)
\ee
Nothing is left, the solution is
\be
h^x_y = -\pi Q^2\bigg( i~ u J_2 (\sqrt{4Q^2 u}) + u Y_2 (\sqrt{4Q^2 u}) \bigg)
\ee
then the spectral function is
\be
\chi_{xyxy}^{T=0}= \frac{l^3}{16 G_5^2} \left(\frac{2\pi T}{1-a/2}\right)^4~ \Im\left[\lim_{u \rightarrow 0} \frac{f(u)}{u}h^x_y {h^x_y}'\right] = \frac{l^3}{16 G_5^2} (w^2-k^2)^2 \theta(w^2-k^2)
\ee
where the theta function comes from the fact $\sqrt{Q^2}$ should be real.

\subsection{Vector mode}
For the vector mode, we have two master variables $\Psi_+, \Psi_-$ and their equations of motions. These can be transformed into Schrodinger form by
\be
\Psi_\pm = \frac{1}{\sqrt{f}} \psi_\pm
\ee
then equations of motion is
\be
{\psi_\pm}'' +V_\pm \psi_\pm =0, \quad V_\pm = \frac{\wn^2-\qn^2 f}{uf^2}\left(1-\frac{a}{2}\right)^2+\frac{1}{4}\frac{{f}'^2}{f^2}-\frac{1}{f}\left(C_\pm - \frac{f'}{u}+\frac{f''}{2}\right).
\ee
The equation of motion for the master fields near the boundary is
\ba
0&=&{\psi}''+\frac{(\wn^2-\qn^2)(1-a/2)^2}{u}\psi \equiv {\psi}''+\frac{Q^2}{u}\psi \no
\psi &\sim& C_1 \sqrt{u} J_1(2\sqrt{Q^2u}) +C_2\sqrt{u} Y_1(2\sqrt{Q^2u})
\ea
And near the horizon,
\ba
0&=&{\psi}''+\frac{1+\wn^2}{4(1-u^2)}\psi \no
\psi &\sim& C_3 (1-u)^{1/2-i \wn/2} + C_4 (1-u)^{1/2+i \wn/2}.
\ea
For large $Q^2$, two WKB solutions are
\be
\psi_1 \sim \frac{1}{\sqrt{\pi p(u)}} ~\mathrm{cos}(S(u)+\phi_1), \quad
\psi_2 \sim \frac{1}{\sqrt{\pi p(u)}} ~\mathrm{sin}(S(u)+\phi_2)
\ee
where
\be
p(u) = \frac{1-\frac{a}{2}}{\sqrt{u} f} \sqrt{\wn^2-\qn^2 f},\quad S(u) = \int^u_0 p(z) dz.
\ee
This WKB solutions should be matched near boundary solutions
\ba
2 \sqrt{u} J_1(2 \sqrt{Q^2 u}) &\sim& \sqrt{u} \sqrt{\frac{1}{\pi \sqrt{Q^2u}}} \mathrm{cos}\left(2 \sqrt{Q^2 u}-\frac{3\pi}{4}\right) \no
2 \sqrt{u} Y_1(2 \sqrt{Q^2 u}) &\sim& \sqrt{u} \sqrt{\frac{1}{\pi \sqrt{Q^2u}}} \mathrm{sin}\left(2 \sqrt{Q^2 u}-\frac{3\pi}{4}\right)
\ea
and physically relevant boundary condition at horizon, infalling condition $C_4$=0,
\be
\Psi =\frac{\psi}{\sqrt{f}}= 2 \sqrt{u} ~C\bigg(i ~ J_1(2 \sqrt{Q^2 u}) + Y_1(2 \sqrt{Q^2 u})\bigg) = - \pi Q \sqrt{u} \bigg(i ~ J_1(2 \sqrt{Q^2 u}) + Y_1(2 \sqrt{Q^2 u})\bigg)
\ee
and by the normalization condition, $\Psi(u=0)=1$ the coefficient C = -$\pi Q/2$. Then the spectral function for vector mode is given as
\ba
\chi_{xx}^{T=0}&=&\frac{l}{4e^2b^2}\Im \frac{C_+ \hat{\Pi}_+^{T=0} - C_- \hat{\Pi}_-^{T=0}}{C_+-C_-} \no
&=& \frac{l}{4e^2b^2} \pi Q^2 = \frac{l}{4e^2} (2\pi T)^2 \pi (\wn^2-\qn^2)\theta(w^2-k^2)
\ea
Note that with Chern-Simons term, there is no difference without CS term for the large frequency spectral function.
\subsection{Light like momenta}
For the light like momenta, w=k, the equation of motion is simplified
 \cite{Atmaja:2008mt} by
\be \label{lightlikeqem}
{\psi_\pm}'' +(\tilde{\wn}^2 H+G_\pm) \psi_\pm =0, ~~\mbox{where} \quad H = \frac{1- f}{uf^2}, \quad G_\pm = \frac{1}{4}\frac{{f}'^2}{f^2}-\frac{1}{f}\left(C_\pm - \frac{f'}{u}+\frac{f''}{2}\right).
\ee
Near the black hole horizon the solution should be infalling, $(1-u)^{-i \frac{\wn}{2}}$. The analytic solution for the equation is not known but we just consider the large $\wn$ limit only to compute asymptotic behavior of spectral function. For large $\wn$, the leading term is $\tilde{\wn}^2$ and we assume the change of the wave function in the domain u $\in(0,1)$ is not large.

For the large $\wn$ limit the first term in eq. (\ref{lightlikeqem}) is dominant. Introduce a new variable $\zeta$
\be
\zeta = \left[\frac{3}{2}\int^u_0 \sqrt{-H(x)}dx\right]^{2/3},
\ee
the wavefunction is reexpressed as
\ba
\psi_\pm &=& \left(\frac{d\zeta}{du}\right)^{-1/2} W = \left(\frac{-H}{\zeta}\right)^{-1/4}W \no
0&=&\frac{d^2 W}{d \zeta^2} -\left( \tilde{\wn}^2\zeta +\gamma\right) W, \quad \mbox{where} ~\gamma = \frac{5}{16 \zeta^2}+\left(\frac{4HH''-5H'^2}{16H^3} +\frac{G}{H}\right)\zeta
\ea
where prime denotes derivative with respect to u, then the solution $\psi_\pm$ is Airy function,
\be
\Psi_\pm = \frac{1}{\sqrt{f}} \psi_\pm  = \frac{1}{\sqrt{f}}\left(\frac{\zeta}{-H}\right)^{1/4} Ai(\tilde{\wn}^{2/3}\zeta(u)) + \cdots = \left(\frac{u \zeta}{f-1}\right)^{1/4} Ai(\tilde{\wn}^{2/3}\zeta(u)) + \cdots
\ee
Two point function for master variables are given as
\be
\Pi_\pm = \lim_{u \rightarrow 0}\frac{\Psi_\pm'}{\Psi_\pm} = \lim_{u \rightarrow 0} \left(\frac{1}{4}\deriv_u \ln\left(\frac{u\zeta}{1-f}\right)+\frac{\deriv_u Ai(\tilde{\wn}^{2/3}\zeta(u)) }{Ai(\tilde{\wn}^{2/3}\zeta(u)) }\right)
\ee
since $f-1=-u^2 (1 + a - a u)$, and
\be
\lim_{u \rightarrow 0}\zeta= \frac{(1+a)^{1/3} \left(i \left(4+3 a+\sqrt{1+a}(4+a)\right)\right)^{2/3} }{\left(\sqrt{1+a}+1\right)^2} u +\cdots .
\ee
The second term in equation is
\be
\lim_{u \rightarrow 0}\deriv_u \ln\left(\frac{u\zeta}{1-f}\right) = \lim_{u \rightarrow 0}\deriv_u \ln\left(\frac{\zeta}{u (1 + a - a u)}\right) \sim \lim_{u \rightarrow 0}\deriv_u \ln\left(\frac{1}{(1 + a )}\right) =0
\ee
\begin{figure}
\begin{center}
    \includegraphics[angle=0, width=0.55 \textwidth]{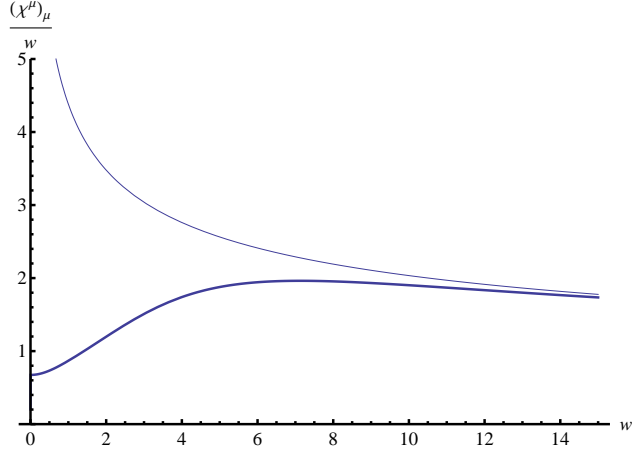}
  \caption{The normalized trace of thermal spectral function $\chi^\mu_\mu/\wn$ with light like momenta. Thick line is for $\bar{\mu}$ =5 and thin line is for zero temperature case. }\label{smeasure}
\end{center}
\end{figure}
So the two point function is
\ba
\Pi_\pm &=& -\mathrm{e}^{i \pi/3}\frac{3^{1/3} (1+a)^{1/3} \Big(4+3 a+\sqrt{1+a} (4+a)\Big)^{2/3} }{\left(1+\sqrt{1+a}\right)^2} \frac{\Gamma\left(\frac{2}{3}\right)}{\Gamma\left(\frac{1}{3}\right)} \tilde{\wn}^{2/3}  \no
\Im \Pi_\pm & = &  \frac{3^{5/6} (1+a)^{1/3} \Big(4+3 a+\sqrt{1+a} (4+a)\Big)^{2/3} }{2\left(1+\sqrt{1+a}\right)^2} \frac{\Gamma(2/3)}{\Gamma(1/3)} \tilde{\wn}^{2/3} \\
\chi_\mu^\mu &=& \frac{l}{4 e^2}\frac{(2\pi T)^2}{(1-a/2)^{4/3}}\frac{3^{5/6} (1+a)^{1/3} \Big(4+3 a+\sqrt{1+a} (4+a)\Big)^{2/3} }{\left(1+\sqrt{1+a}\right)^2} \frac{\Gamma(2/3)}{\Gamma(1/3)} \wn^{2/3} \nonumber
\ea

\section{Small $\wn$ spectral funcions} \label{smallwsp}
To check the consistency of numerical calculation, it is good to compare both large frequency and low frequency result with numerical computation. In the low frequency limit, authors \cite{RNtvmode} did the hydrodynamic analysis and it also gives us the low frequency spectral function. For the tensor mode,
\ba
\lim_{\wn \rightarrow 0}\Im~ G_{xyxy} &=& \frac{l^3}{16G_5^2} \frac{(2\pi T)^4}{(1-a/2)^3}\wn \no
\lim_{\wn \rightarrow 0}\frac{\chi_{xyxy}}{\wn} &=& \frac{l^3}{16G_5^2} \frac{(2\pi T)^4}{(1-a/2)^3}.
\ea
And for the vector mode $G_{xx}$,
\ba
\lim_{\wn \rightarrow 0}\Im~ G_{xx} &=& \frac{l}{4e^2} \bigg[\frac{3a}{(1+a)b^2}\frac{Dk^2w}{D^2k^4+w^2}+\frac{2(1-a/2)^2}{(1+a)^2b}w \bigg] \no
\lim_{\wn \rightarrow 0} \frac{\chi_{xx}}{\wn} &=& \frac{l}{4e^2} \frac{(2\pi T)^2}{(1-a/2)^2}\bigg[\frac{3a}{1+a}\frac{D \qn k}{D^2\qn^2 k^2+\wn^2}+2\frac{(1-a/2)^3}{(1+a)^2} \bigg] \no
\lim_{\wn \rightarrow 0}\lim_{\qn \rightarrow 0}\frac{\chi_{xx}}{\wn} &=& (2\pi T)^2 \frac{l}{4e^2}\frac{2(1-a/2)}{(1+a)^2},
\ea
where D = $\frac{b}{2(1+a)}$. For the $G_{xtxt}$ in small $\wn$, $\qn$ limit,
\ba
\Im~ G_{xtxt} &=& \frac{l^3}{16G_5^2} \frac{1}{b^3} \frac{w k^2}{D^2 k^4+w^2} \no
\chi_{xtxt} &=& \frac{l^3}{16G_5^2}  \frac{(2 \pi T)^4}{(1-a/2)^3} \frac{\wn \qn^2}{D^2\qn^2 k^2+\wn^2}
\ea

\section{Boundary action}\label{bdryaction}
In this section, we will briefly mention how to get rid of the divergences from the on-shell gravity action. The gauge/gravity correspondence tells us that the generating functional of the gauge theory is identified with the generating functional of the AdS gravity. As we have seen in section \ref{recipeforgreenf}, from the generating functional we get the two point functions of the boundary theory G$^{ret}_{\mu\nu}$ or G$^{ret}_{\alpha\beta,\mu\nu}$. The generating functional has some divergences which could be safely removed by adding counter terms, so called holographic renormalization. At the boundary u=0, there are two types of divergences 1/u and Log u.

The original action eq.(\ref{RNaction1}) has Einstein-Hilbert, Maxwell and Gibbons-Hawking term and to remove the divergences we need the following counter term \cite{Skenderis:2002wp} for the regularized action at the boundary,
\ba \label{ct}
S_{\rm ct} &=& S_{\rm ct, \ gravity}+S_{\rm ct, \ gauge} \no
&=&\frac{1}{16 G_5^2} \! \int \! \dd^4x \sqrt{-g^{(4)}}
\left( \frac{3}{l} - \frac{l}{4} K \right)+ \frac{l}{8e^2} \log u \! \int \! \dd^4x \sqrt{-g^{(4)}} {\cal F}_{mn} {\cal F}^{mn}
\ea
where K is the curvature on the boundary. $S_{\rm ct,\ gravity}$ is given in \cite{Balasubramanian:1999re}. On the other hand, $S_{\rm ct,\ gauge}$ is obtained to cancel the logarithmic divergence
coming from gauge field fluctuations. The boundary action for the perturbation in quadratic order derived from (\ref{RNaction1}) is
\begin{eqnarray}
\label{sgra}
S^{(0)} &=&
\frac{l^3}{32 b^4 G_5^2} \!\int\!\frac{\dd^2k}{(2\pi)^2}\frac{1}{u^2}
\Bigg[
\frac{u f'}{f}   (h^x_t)^2
+   h^x_t     \Big(h^x_t - 3 u {h^x_t}'\Big)
- f h^x_z     \Big(h^x_z - 3 u {h^x_z}'\Big)
\no
&& \hspace*{35mm} + 3a B_x \Big(h^x_t - f {B_x}'\Big) \Bigg]\ .
\end{eqnarray}
The Gibbons-Hawking term and the counter term (\ref{ct}) are
\begin{eqnarray}
\label{sgh}
S^{(0)}_{\rm GH}
&=&
\frac{l^3}{32 b^4 G_5^2} \frac{1}{u^2} \!\int\!\frac{\dd^2k}{(2\pi)^2}
\Bigg\{ - \frac{u f'}{f} (h^i_t)^2 - 4 (h^i_t)^2
- u f' (h^i_z)^2 + 4 u h^i_t {h^i_t}'  \nonumber \\
&& \hspace*{38mm}
+ 4 f \Big( (h^i_z)^2 - u h^i_z(h^i_z)' \Big)  \Bigg\} \ ,   \\
\label{sct}
S^{(0)}_{\rm ct}
&=&
\frac{3 l^3}{32 b^4 G_5^2} \frac{1}{u^2 \sqrt{f}}\!\int\!\frac{\dd^2k}{(2\pi)^2}
\Big( (h^i_t)^2 - f (h^i_z)^2 +ab^2k^2 u^2f(u)\log u \Big(B_x^2\Big)\Big) \ .
\end{eqnarray}
Then, the regularized boundary action is given as
\ba \label{surface}
S_{\rm bdry}
&=&
\lim_{u \to 0}(S^{(0)} + S^{(0)}_{\rm GH} + S^{(0)}_{\rm ct} ) \no
&=& \lim_{u \to 0}\frac{l^3}{32 b^4 G_5^2} \!\int\!\frac{\dd^2k}{(2\pi)^2}
\Bigg\{
\frac{1}{u}\Big(h^x_t(-k){h^x_t}'(k)-h^x_z(-k){h^x_z}'(k)\Big) \no
&& + 3a B_x(-k)\Big( h^x_t(k)-{B_x}'(k)\Big) +3ab^2k^2 \log u \Big(B_x(-k)B_x(k)\Big)
\Bigg\} .
\ea

\end{document}